\documentclass[12pt]{iopart}
\usepackage[latin1]{inputenc}
\usepackage{graphicx}
\usepackage{setstack}
\usepackage{subfigure}
\usepackage{iopams} 
\usepackage{dcolumn}
\newcolumntype{d}[1]{D{.}{.}{#1}}

\begin{document}
\title[Spectral and fluence constraints using optimal control theory]{Tailoring laser pulses with spectral and fluence constraints using optimal control theory}
\author{J. Werschnik and E.K.U. Gross}
\address{Institut für Theoretische Physik, Freie Universität  Berlin, Arnimallee 14, 14195 Berlin, Germany}
\date{\today}
\ead{jan.werschnik@physik.fu-berlin.de}
\begin{abstract} 
Within the framework of optimal control theory we develop a simple iterative scheme to determine optimal laser pulses with spectral and fluence constraints. The algorithm is applied to a one-dimensional asymmetric double well where the control target is to transfer a particle from the ground state, located in the left well, to the first excited state, located in the right well.
Extremely high occupations of the first excited state are obtained for a variety of spectral and/or energetic constraints. Even for the extreme case where no resonance frequency is allowed in the pulse the algorithm achieves an occupation of almost $100\%$.
\end{abstract}

\submitto{\JOB}
\section{Introduction}
\label{sec:ntro}
%
In the last fifteen years, the control of quantum mechanical systems via light pulses has experienced/seen considerable progress, both on the experimental and  on the theoretical side. Quantum control experiments have been pushed forward by the improvement of laser pulse shaping \cite{WLPW92,BKPSG2005,SWYYH2005} and the implementation of closed-loop learning (CLL) techniques \cite{JR92}. Experiments using CLL delivered highly encouraging results, ranging from the control of chemical reactions \cite{B97,A98,LMR2001,BDNG2001,HWCZM2002,D2003,BDKG2003,VKNNG2005} to the control of high-harmonic generation \cite{B2000,PWWSG2005}.

On the theoretical side, the most important contributions have been the introduction of optimal-control theory \cite{K89,PDR88,HTC83} and the continuous development of  rapidly converging iteration schemes \cite{ZBR98,ZR98,MT2003} to calculate optimal laser pulses. Recently, some of these schemes have been generalized to include dissipation \cite{OZR99}, to account for multiple objectives \cite{O2001} and to deal with time-dependent control targets \cite{OTR2004,KM2004,SWG2005}. 

Most fruitful are investigations where theory and experiment come together: The theoretical analysis of laser pulses can be useful, and sometimes is essential, in deciphering the pulse shapes obtained from experiment \cite{B2004}. To speed up the convergence process of the experimental learning cycle, calculated pulses can be used either to provide an initial guess or to reduce the gigantic search space by determining the most important shape parameters \cite{ZFKM2001}.
Besides these direct applications, theory and computer simulations make it possible to explore the feasibility of future experiments and help to determine the requirements on the laser system and the pulse shaping device.
Computer simulations can also help in understanding new ultrafast transition processes in laser-assisted chemistry \cite{DKMS98,GHV2003} and in developing new implementations for the quantum computer \cite{PK2002,TR2004}.

For all these applications it is extremely important that the computational schemes are able to include experimental constraints, such as limitations on the spectral bandwidth and on the laser fluence, i.e. the time-integrated intensity. As discussed in \ref{app_post}, a pulse from an unconstrained optimization will perform much worse, if the constraint is applied ``brute force'' after the optimization than a pulse coming from a scheme where the same constraints are built in.

So far, only few attempts have been made to take restrictions of this kind into account. In reference \cite{K89} a scheme to calculate the pulse for a given fluence is shown. However, it does not make use of the immediate feedback introduced in \cite{ZBR98} and suffers from a rather unstable convergence. A constraint on the spectrum is considered in \cite{GNR92} for a steepest descent method which, in the quantum control context, is found to suffer from poor convergence and a strong dependence on the initial pulse \cite{SST98}. An elegant way to restrict the spectrum has been presented by the authors of reference \cite{HMV2001}. This scheme preserves the rapid and monotonic convergence behavior of the underlying scheme \cite{ZBR98} by projecting out undesired parts of the time-dependent wave-function, which are responsible for the unwanted spectral components. However, this method is not sufficiently general and does not easily allow for an additional fluence constraint (as it keeps the unphysical penalty factor). 

The scheme presented in the following allows one to incorporate fluence and/or spectral constraints in the optimization and it shows very good convergence, when applied to a 1D model, although a proof of monotonic convergence similar to reference \cite{ZR98} does not go through here. Further more, the scheme is very simple to implement. 

An introduction to optimal control theory is given in \sref{sec:theory}. We then explain our schemes in \sref{algorithm}. In \sref{sec:compdetails} we present a test system and discuss the numerical details. The results from applying our algorithms to this system are analyzed in \sref{sec:results}.

\section{Optimal Control Theory}
\label{sec:theory}
%
In this section we sketch the basics of optimal control theory applied to quantum mechanics.
We consider an electron in an external potential $V({\bf r})$ under the influence of a laser field propagating in $z$-direction. Given an initial state $\Psi({ \bf r },0)=\phi({ \bf r })$ the time evolution of the electron is described by the  time-dependent Schr\"odinger equation with the laser field modeled in dipole approximation (length gauge)
\begin{eqnarray}
i\frac{\partial}{\partial t}\Psi({ \bf r },t)&=&\widehat{H}\Psi({ \bf r },t),\label{1SE}\\
\widehat{H}&=&\widehat{H}_0-\hat{{\bmu}}{\bepsilon}(t),\\
\widehat{H}_0&=&\widehat{T}+\widehat{V},
\end{eqnarray}
(atomic units are used throughout: $\hbar = m =e =1 $). Here, $\hat{{\bmu}}=(\hat{\mu}_x,\hat{\mu}_y)  $ is the dipole operator, and ${\bepsilon}(t)=(\epsilon_x(t),\epsilon_y(t))$ is the time-dependent electric field. The kinetic energy operator is $\widehat{T}=-\frac{\nabla^2}{2}$.\\
Our goal is to control the time evolution of the electron by the external field in such a way that the expectation value of the target operator $\widehat{O}$ is maximized with respect to the wave function at the end of the pulse $\Psi({\bf r},T)$.  Mathematically, this goal corresponds to maximizing the functional \cite{K89, ZBR98,ZR98}:
\begin{eqnarray} \label{eq:J1}
J_1[\Psi]&=& \langle \Psi(T)|\widehat{O}|\Psi(T)\rangle.
\end{eqnarray} 
Usually $\widehat{O}$ is assumed to be positive-semidefinite which guarantees monotonic convergence of the schemes discussed in references \cite{ZBR98,ZR98,MT2003,OTR2004}. A few examples will be discussed at the end of this section.\\
%
%
The functional $J_1[\Psi]$ will be maximized subject to a number of physical constraints.
The idea is to cast also these constraints in a suitable functional form and then calculate the total variation. Subsequently, we set the total variation to zero and find a set of coupled partial differential equations \cite{K89,PDR88}. The solution of these equations will yield the desired laser field ${ \bepsilon }(t)$.\\
In more detail: Optimizing $J_1$ may possibly lead to fields with very high, or even infinite energy. In order to avoid these strong fields, we include an additional term in the functional which penalizes the fluence of the field. This can be done for each polarization direction separately:
\begin{eqnarray}
J_2[{ \bepsilon }] &=& - \sum_j  \int_0^T \!\! dt \,\, \alpha_j {  \epsilon_j }^2(t) \qquad j=x,y,  
\end{eqnarray} 
where $\alpha_j$ is a penalty factor that has to be chosen. It balances the optimization between increasing the yield and restricting the energy to achieve the maximal value for the combined functional $J_1 + J_2$. Note, the penalty factor $\alpha_j$ can be made time-dependent to restrict the laser pulse to a certain shape \cite{SV99}.\\

The constraint on the laser fluence can be expressed also in another way:
\begin{eqnarray}
\label{eq:fluence}
\tilde{J}_2[{ \bepsilon }] &=& - \sum_j  \alpha_j \left[ \int_0^T \!\! dt \,\, {  \epsilon_j }^2(t) -E_{0_j} \right]. 
\end{eqnarray}
Here, $\alpha_j$ is a (time-independent) Lagrange multiplier. Instead of specifying $\alpha_j$ we have to prescribe specific values, $E_{0_j}$, for the components $E_{0_x}$ and $E_{0_y}$ of the laser fluence. Hence, this approach requires two Lagrange multipliers $\alpha_x$ and $\alpha_y$. 

The constraint that the electronic wave function has to fulfill the time-dependent Schr\"odinger equation is expressed by
\begin{eqnarray}
  J_3[{ \bepsilon },\Psi,\chi]&=&
    - 2 \Im \int_{0}^{T}\!\!  dt \,\, \left\langle\chi(t) \left| \left(i\partial_t
    -\widehat{H}\right) \right| \Psi(t)\right\rangle.
\end{eqnarray} 
with a Lagrange multiplier $\chi({ \bf r },t)$. $\Psi({ \bf r },t)$ is the wave function driven by the laser field ${ \bepsilon }(t)$.\\
The Lagrange functional has the form
\begin{eqnarray}
J[\chi,\Psi,{ \bepsilon }] = J_1[\Psi] + J_2[{ \bepsilon }] + J_3[\chi,\Psi,{ \bepsilon }].
\end{eqnarray} 
Setting the variations of the functional with respect to $\chi$, $\Psi$, and ${ \bepsilon }$ independently to zero yields
\begin{eqnarray}
 \alpha_j \epsilon_j(t) &=& -\Im\langle\chi(t)|\hat{\mu}_j|\Psi(t)\rangle, \label{eq:field} \qquad j=x,y\\
0 &=& \left( i \partial_t - \widehat{H} \right) \Psi({ \bf r },t), \qquad 
\Psi({ \bf r },0) = \phi({ \bf r }),\label{eq:SE}\\
&&\left(i\partial_t - \widehat{H}\right)\chi({ \bf r },t) =\nonumber\\
\label{eq:SE1}
\lo &&i\left(\chi({ \bf r },t)-\widehat{O} \Psi({ \bf r },t)\right)\delta(t-T).
\end{eqnarray}
\Eref{eq:field} determines the field from the wave function $\Psi({ \bf r },t)$ and the Lagrange multiplier $\chi({ \bf r },t)$.\\
\Eref{eq:SE} is a time-dependent Schr\"odinger equation for $\Psi({ \bf r },t)$ starting from a given initial state $\phi({ \bf r })$ and driven by the field ${ \bepsilon }(t)$. 
If we require the Lagrange multiplier $\chi({ \bf r },t)$ to be continuous, we can solve the following two equations instead of \eref{eq:SE1}:
\begin{eqnarray}
\label{eq:SE2}
 \left(i\partial_t - \widehat{H}\right)\chi({ \bf r },t)&=&0,\\
\label{eq:SE3}
\chi({ \bf r },T) &=&\widehat{O} \Psi({ \bf r },T),
\end{eqnarray}
To show this we integrate over \eref{eq:SE}
\begin{eqnarray}
\nonumber
&& \lim_{\kappa\to 0}\int_{T-\kappa}^{T+\kappa}\!\!\!\!\!\!dt\left[\left(i\partial_t - \widehat{H}\right)\chi({ \bf r },t) \right]\\
\label{eq:proof1}
&=&\lim_{\kappa\to 0}\int_{T-\kappa}^{T+\kappa}\!\!\!\!\!\!dt\; i\left(\chi({ \bf r },t)-\widehat{O}(t) \Psi({ \bf r },t)\right)\delta(t-T).
\end{eqnarray}
The left-hand side of \eref{eq:proof1} vanishes because the integrand is a continuous function. It follows that also the right-hand side must vanish, which implies \eref{eq:SE3}. From equations (\ref{eq:SE3}) and (\ref{eq:SE1}) then follows equation \eref{eq:SE2}.\\
Hence, the Lagrange multiplier $\chi({\bf r} ,t)$ satisfies a time-dependent Schr\"odinger equation with an initial condition at $t=T$. 
The set of equations that we need to solve is now complete:  (\ref{eq:field}),  (\ref{eq:SE}), (\ref{eq:SE2}) and (\ref{eq:SE3}). 
If we use $\tilde{J}_2$ instead of $J_2$ we also have to perform a variation with respect to $\alpha_j$ which simply yields the restriction:
\begin{eqnarray}
\int_0^T \!\! dt \,\, \epsilon_j^2(t) = E_{0_j}.
\end{eqnarray}
To find an optimal field ${ \bepsilon }(t)$ from these equations we use an iterative algorithm which is discussed in the next section.\\ 

We conclude this section with a discussion of  the target operator $\widehat{O}$. Basically, there exist two classes of target operators. Namely, operators that are non-local, e.g. projection operators and operators that are local (multiplicative), like the density operator. 
If we want to maximize the occupation of a given target state $|\Phi_f \rangle$ at the end of the laser pulse, we choose \cite{K89,ZBR98},
\begin{eqnarray}
\label{eq:target_proj}
\widehat{O} = | \Phi_f \rangle \langle \Phi_f |.
\end{eqnarray}
This scheme can be extended to achieve multiple goals, i.e. to have different states populated at the end of the pulse. In that case one uses,
\begin{eqnarray}
\widehat{O} = \sum_k \beta_k | \varphi_k \rangle \langle \varphi_k |.
\end{eqnarray}
The factors $\beta_k$ allow for the possibility to ``fine-tune'' the target occupations among each other (multi-objective optimization), i.e. to balance between the importance of the individual targets. For example, if we choose $\beta_n$ negative the optimization will avoid the occupation of the state~$| \varphi_n \rangle$.

Note, that this kind of multi-objective optimization is different from the target to reproduce a (coherent) superposition of field free eigenstates of the Hamiltonian $H_0$ given by 
\begin{eqnarray}
\nonumber
\widehat{O} &=& | \Phi_f \rangle \langle \Phi_f |, \\
| \Phi_f \rangle &=& \sum_k c_k | \varphi_k \rangle .
\end{eqnarray}

Using as target operator the projection operator (\ref{eq:target_proj}) leaves the freedom of a purely time-dependent phase factor for the wave function $\Psi(T)$. It is possible to fix the phase with the following functional:
\begin{eqnarray}
\min \| \Psi(T) - \Phi_f \|^2 = 2\, \left( 1-\Re \langle \Psi(T) | \Phi_f \rangle \right)\\
\Rightarrow \tilde{J}_1 = \max \Re \langle \Psi(T) | \Phi_f \rangle,
\end{eqnarray}
where we have assumed normalization: $\langle \Psi(T) |\Psi(T) \rangle=\langle \Phi_f |\Phi_f \rangle=1$.

The target operator may also be local \cite{ZR98}. If we choose $\widehat{O} = \delta({ \bf r }-{ \bf r }_0)$ (the density operator), we maximize the probability density in ${ \bf r }_0$ at $t=T$:
\begin{eqnarray}
\label{eq:loc_op}
J_1 = \langle \Psi(T)| \widehat{O} |\Psi(T) \rangle = n({ \bf r }_0,T). 
\end{eqnarray}
For this control target, the optimization process will try to concentrate the density in the point ${ \bf r }_0$ at the end of the pulse \cite{janphd}.
Numerically, the $ \delta$-function can be approximated by a sharp Gaussian function. 
%

\section{Algorithm}
\label{algorithm}
%
In this section we present iterative schemes for the optimization of laser fields under additional constraints on the fluence and/or on the spectral distribution.
\subsection{Fluence constraint}
%
We first describe the algorithm which yields an optimized laser pulse producing an assigned value of $E_{0_j}$ (for each polarization direction $j=x,y$, cf. equation \eref{eq:fluence}). The set of coupled equations to be solved is given by equations (\ref{eq:field}),  (\ref{eq:SE}), (\ref{eq:SE2}) and (\ref{eq:SE3}).
The scheme below shows the order in which these equations are solved in the $k$th iterative step.
\begin{equation}
\label{eq:scheme}
\begin{array}{l c c l c c c c }
  {\mbox{k-th step:}} \,\, & \Psi^{(k)}(0) & \overset{{\bepsilon}^{(k)}(t)}{\longrightarrow} &
 \Psi^{(k)}(T) &  &  &  &\\
          & & & \left[ \Psi^{(k)}(T) \right. & \overset{{\bepsilon}^{(k)}(t)}{\longrightarrow} &
 \left. \Psi^{(k)}(0) \right] & & \\
                   & & & \chi^{(k)}(T) & \overset{\widetilde{{\bepsilon}}^{(k)}(t) 
}{\longrightarrow} & \chi^{(k)}(0), & &
\end{array}
\end{equation}
with the laser fields $\bepsilon^{(k)}(t),\tilde{\bepsilon}^{(k)}(t)$ given by
\begin{eqnarray}
\label{eq:feld1}
\widetilde{\epsilon}_j^{(k)}(t) &=& -  \frac{1}{\alpha_j^{(k)}}\Im\langle\chi^{(k)}(t)|\hat{\mu}_j|\Psi^{(k)}(t)\rangle,\\
\label{eq:feld2}
\epsilon_j^{(k+1)}(t) &=& \frac{\alpha_j^{(k)} }{\alpha_j^{(k+1)}} \widetilde{\epsilon}_j^{(k)}(t), \qquad \qquad j=x,y,
\end{eqnarray}
where the Lagrange multiplier $\alpha^{(k+1)}_j$ is defined by:
\begin{eqnarray}
\label{eq:alpha1}
\alpha_j^{(k+1)} = \sqrt{\frac{\int_0^T \!\! dt  \left[ \alpha_j^{(k)} \widetilde{\epsilon}_j^{(k)}(t) \right]^2 }{E_{0_j}}}.
\end{eqnarray}
The initial conditions in every iteration step are
\begin{eqnarray}
\label{eq:initial_psi}
  \Psi({\bf r},0)&=& \phi({\bf r}),\\
\label{eq:initial_chi}
  \chi({\bf r},T) &=& \widehat{O} \Psi({\bf r},T).
\end{eqnarray}

The scheme starts with the propagation of $\Psi^{(0)}({\bf r},t)$ forward in time using the laser field $\epsilon^{(0)}(t)$ which has to be guessed. The result of the propagation is the wave-function $\Psi^{(0)}({ \bf r },T)$ which is now used to calculate $\chi^{(0)}({\bf r},T)$ by applying the target operator (\ref{eq:initial_chi}). We continue with propagating  $\chi^{(0)}({ \bf r },t)$ backwards in time using the laser field $\widetilde{\epsilon}^{(0)}(t)$ (\ref{eq:feld1}). To solve equation \eref{eq:feld1} we have to know both wave functions $\Psi^{(0)}({ \bf r },t)$ and $\chi^{(0)}({ \bf r },t)$ at the same time $t$, which makes it necessary to either store the whole time-dependent wave function $\Psi^{(0)}({ \bf r },t)$ or propagate it backwards with the previous laser field $\epsilon^{(0)}(t)$. The version of the algorithm that avoids storage is indicated by the brackets in the scheme (\ref{eq:scheme}). Besides that, it is necessary to provide an initial value for $\alpha_j^{(0)}$ which we choose to be:
\begin{eqnarray}
\nonumber
\alpha_j^{(0)} = \sqrt{\frac{\int_0^T \!\! dt \,\,  \left[\epsilon_j^{(0)}(t)\right]^2}{E_{0_j}}}.
\end{eqnarray}
The result of the backward propagation $\chi^{(0)}({ \bf r },t)$ is the laser field $\widetilde{\epsilon}^{(0)}(t)$ which we now re-scale to the right value (\ref{eq:feld2}) yielding $\epsilon^{(1)}(t)$. This completes the first step. The second $(k=1)$ or, in general, the $k$th iteration repeats the described procedure starting again from the initial state $\Psi^{(k)}({ \bf r },t)=\phi({\bf r})$ and applying the rescaled field $\epsilon^{(k)}(t)$.

The scheme described above has some aspects in common with the techniques described in reference \cite{K89} and reference \cite{ZBR98}:
The basic idea of incorporating fluence constraints in the optimization algorithm was given in reference \cite{K89}. However, the authors do not make use of the immediate feedback (cf. equation \eref{eq:feld1}), i.e. the backward propagation is accomplished by updating $\chi({ \bf r },t)$ and $\epsilon(t)$ in a self-consistent way, which was suggested in reference \cite{ZBR98}. On the other hand, the technique presented by the authors of reference \cite{ZBR98} does not allow to build in fluence constraints, since $\alpha_j$ is not a Lagrange multiplier in their case.
Roughly speaking, the technique presented above is a combination of both approaches. 
%
\subsection{Spectral constraint}
%
The algorithm with built in spectral restrictions is similar to the one presented above with two important differences: The factor $\alpha_j$ is a penalty factor. It has to be specified from the start and remains unchanged during the optimization. Second, the update of the field $\epsilon_j^{(k+1)}(t)$ in equation \eref{eq:feld2} has to be replaced by:
\begin{eqnarray}
\label{eq:feld2_filter}
\epsilon_j^{(k+1)}(t) &=&  \mathcal{F} \left[ f_j(\omega) \times \mathcal{F}\left[\widetilde{\epsilon}_j^{(k)}(t)\right]\right]  \qquad j=x,y,
\end{eqnarray}
where the symbol $\mathcal{F}$ indicates a Fourier-transform.
The spectral constraint is formulated in terms of a filter function $f_j(\omega)$. Since $\epsilon_j(t)$ is real valued we have to make sure that $f_j(\omega) = f_j(-\omega)$. For example, the filter function could be chosen to be: 
\begin{eqnarray}
\label{eq:filter1}
f_j(\omega) =  \exp[-\gamma (\omega-\omega_0)^2] + \exp[-\gamma (\omega+\omega_0)^2],
\end{eqnarray}
so that only the frequency components around the center, $\pm \omega_0$, of the Gaussians are allowed in the pulse. If one uses instead:
\begin{eqnarray}
\label{eq:filter2}
\tilde{f}_j(\omega) = 1- \left(\exp[-\gamma (\omega-\omega_0)^2] + \exp[-\gamma (\omega+\omega_0)^2]\right),
\end{eqnarray}
one would allow every spectral component in the laser field except the components around $\pm \omega_0$.\\
%
%
%
\subsection{Spectral and fluence constraint}
Finally, we note that both schemes can be combined. This combination makes it possible to incorporate even more realistic experimental constraints in computational pulse optimizations. 
This is achieved by the scheme (\ref{eq:scheme}), the equation (\ref{eq:feld1}) and:
\begin{eqnarray}
\label{eq:feld2_filter_fixed}
\bar{\epsilon}_j^{(k)}(t) &=&   \mathcal{F} \left[ f_j(\omega) \mathcal{F}\left[\widetilde{\epsilon}_j^{(k)}(t)\right]\right] \qquad j=x,y,\\
\label{eq:feld2_filter_fixed2}
\epsilon_j^{(k+1)}(t) &=&  \frac{\alpha_j^{(k)}}{\alpha_j^{(k+1)}} \bar{\epsilon}_j^{(k)}(t),
\end{eqnarray}
where $\alpha_j^{(k+1)} $ is evaluated with the filtered field $\bar{\epsilon}_j^{(k)}(t)$
\begin{eqnarray}
\label{eq:alpha_filterfixed}
\alpha_j^{(k+1)} = \sqrt{\frac{\int_0^T \!\! dt  \left[ \alpha_j^{(k)} \bar{\epsilon}_j^{(k)}(t) \right]^2 }{E_{0_j}}},
\end{eqnarray}
to yield the right value for  $E_{0_j}$. The total spectral power is related to the fluence Parseval's theorem:
\begin{eqnarray} 
\nonumber
E_{0_j} = \int_{-\infty}^{+\infty} \!\! dt \,\, \theta(t) \theta(T-t) \left[\epsilon_j(t)\right]^2 = \frac{1}{2 \pi} \int_{-\infty}^{+\infty} \!\! dt \,\, \left| \epsilon_j(\omega)\right|^2 
\end{eqnarray}

In this combined form we first apply the filter function to the laser field (\ref{eq:feld2_filter_fixed}) then we rescale the field to yield the right value for $E_{0_j}$ (see equation \eref{eq:feld2_filter_fixed2}).

We conclude the section with a few remarks:
\begin{itemize}
\item
For each polarization direction one can specify a separate filter or fluence.
\item
The convergence proofs of references \cite{ZBR98,ZR98} do not go through in our case. This is due to the changing value for $\alpha_j^{(k)}$ and, in the case of spectral constraints, due to the modified field (\ref{eq:feld2_filter}). However, as will be shown in \sref{sec:results}, we still find a very good convergence of the presented algorithms in the numerical examples. Even for the ``brute-force'' spectral filter we find a satisfying convergence behavior (unless too many essential features of the pulse are suppressed by the function $f_j(\omega)$).
\item 
Since we do not expect a monotonic convergence we have to add some additional intelligence to the algorithm, e.g. we store the field which produces the pulse with the highest yield and consider this field as the result of the optimization.
\end{itemize}

\section{Computational details and model system}
\label{sec:compdetails}
\begin{figure}[!h]
\centering
\includegraphics*[width=.43\textwidth]{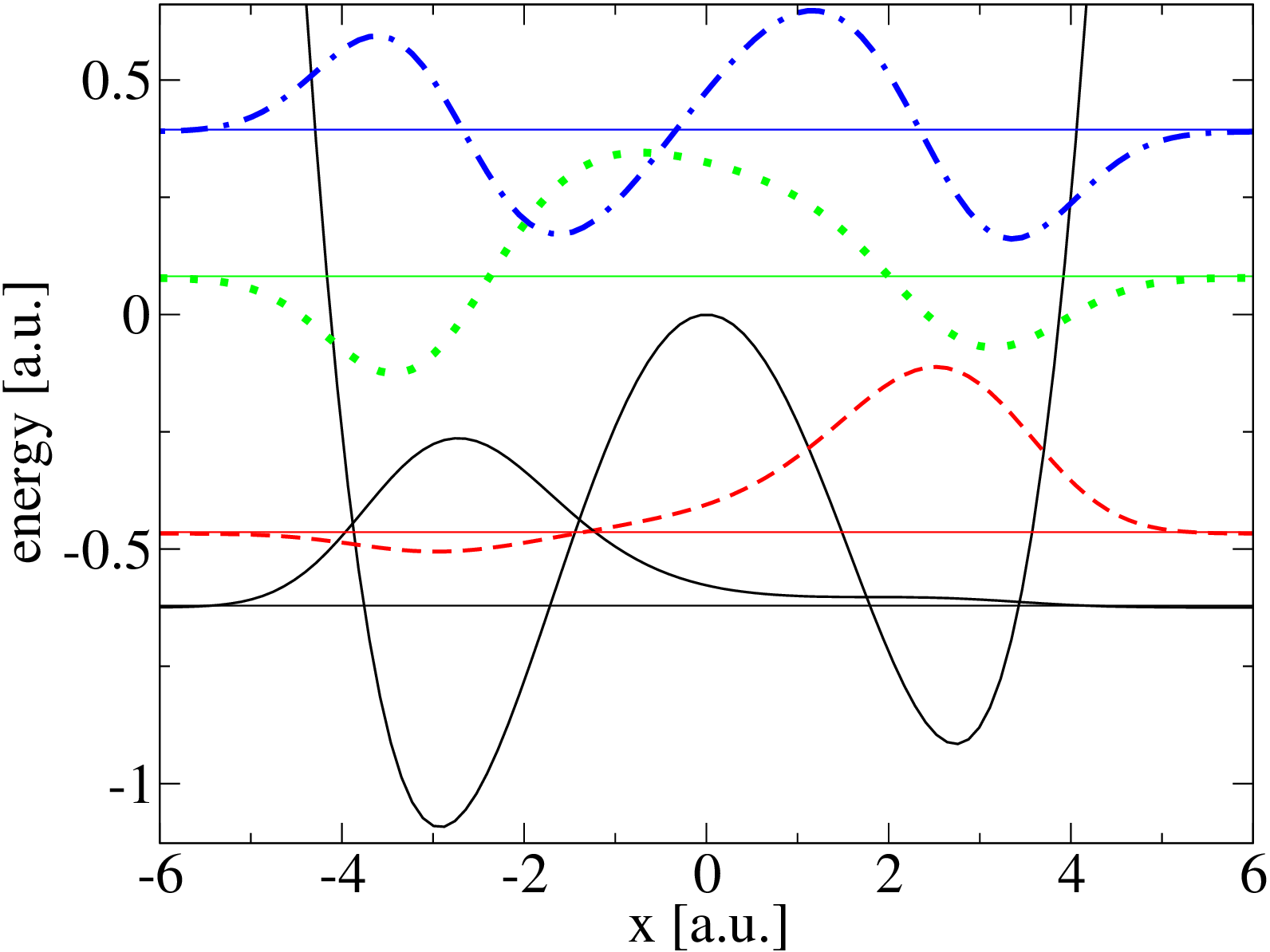}       

\caption{The plot shows the model potential with the ground-state (\full), the first excited state (\dashed), the second excited state (\dotted) and the third excited state (\chain). Each state is shifted according to its eigenvalue.} 
\label{fig:potential_states}
\end{figure}

%
We choose a one-dimensional asymmetric double well to test our algorithms. The double well is similar to reference \cite{GH98} but has an additional  cubic term:
\begin{eqnarray}
V(x) = \frac{w_0^4}{64 B} x^4 -  \frac{\omega_0^2}{4} x^2 + \beta x^3 ,
\end{eqnarray} 
with $\omega_0$ corresponding to the classical frequency at the bottom of the well and the parameter $B$ adjusting the barrier height. The number of pairs of states below the barrier is approximately $B/\omega_0$. Here, we choose $B=\omega_0=1.0$ and $\beta=1/256$ which leads to two states below the barrier, as shown in \fref{fig:potential_states}.
%
In order to analyze the laser pulses from the optimization runs we calculate the excitation energies (see \tref{tab:xenergies}) and dipole moments (see \tref{tab:dipole_moments}) of the system by propagating in imaginary time.

\begin{table}[h]
\begin{indented}
\item[]
\caption{\label{tab:xenergies}Excitation energies in atomic units~[a.u.] for the 1D asymmetric double well, calculated by imaginary time propagation.}
\lineup
\begin{tabular}{@{}lllll}
\br  
 & $|0\rangle$        & $|1\rangle$         & $|2\rangle$        & $|3\rangle$      \\
\mr
 $|0\rangle$   & 0.       &           &          &         \\
 $|1\rangle$   & 0.1568  &  0.       &          &         \\
 $|2\rangle$   & 0.7022  &  0.5454  & 0.       &         \\
 $|3\rangle$   & 1.0147  &  0.8580  & 0.3125  & 0.      \\
 $|4\rangle$   & 1.5294  &  1.3726  & 0.8273  & 0.5147 \\
\br
\end{tabular}
\end{indented}
\end{table}
%
\begin{table}[h]
\begin{indented}
\item[]
\caption{\label{tab:dipole_moments}Dipole matrix elements for the 1D asymmetric double well, calculated by imaginary time propagation.}
\begin{tabular}{@{}ld{4}d{4}d{4}d{4}d{4}}
\br
    &  |0\rangle       & |1\rangle        & |2\rangle        &  |3\rangle       &  |4\rangle  \\
\mr
  $|0\rangle$   & -2.5676  &         &         &         &        \\
  $|1\rangle$   &  0.3921  &  2.3242 &         &         &        \\
  $|2\rangle$   &  0.6382  & -0.7037 & -0.5988 &         &        \\
  $|3\rangle$   & -0.3865  & -0.4630 &  1.7051 &  0.1958 &        \\
  $|4\rangle$   & -0.1414  &  0.2118 &  0.1593 & -1.7862 & -0.0939 \\
\br
\end{tabular}
\end{indented}
\end{table}
%
%

The time-dependent Schr\"odinger equation for the 1D double well is solved on a  grid, where the infinitesimal time-evolution operator is approximated by the 2nd-order split-operator (SPO) technique \cite{FMF76}:
\begin{eqnarray} 
\nonumber
\widehat{U}_{t}^{t+\Delta t}&=&\mathcal{T}\exp\left(-i \int_{t}^{t+\Delta t} \!\! dt' \,\, \widehat{H}(t')\right)\\
\nonumber
\label{eq:spo2nd}
    & \approx  & \exp(-\frac{i}{2}\, \hat {T}\,\Delta t) \exp(-i\, \hat {V}(t)\,\Delta t)\times 
    \exp(-\frac{i}{2}\, \hat {T}\,\Delta t) + O(\Delta t^3).
\end{eqnarray}
Following the scheme described in \sref{algorithm}, one needs three propagations per iteration (if we want to avoid storing the wave function). Within the 2nd order split-operator scheme each time step requires 4 Fast Fourier Transforms (FFT) \cite{FFTW98} for the backward propagations, because we have to know the wave-function and the Lagrange multiplier in real space at every time-step to be able to evaluate the field from equation \eref{eq:field}. For the forward propagation we only need 2 FFTs. This sums up to $10$ FFTs per time step and iteration. \\
\begin{table}[h]
\begin{indented}
\item[]
\caption{\label{tab:parameter}In ordinary runs the listed numerical parameters (given in atomic units) were employed. For the scans we have used a coarser grid in space and time, as indicated in the second column.}
\begin{tabular}{@{}ld{4}d{4}l}
\br
parameter     & \mbox{single run} & \mbox{scan} & \\
\mr
$T$           & 400.0 & 400.0  & pulse length  \\   
$x_{\mathrm{max}}$  & 30.0 & 20.0 & grid size \\
$dx$          & 0.1172  & 0.1563 & grid spacing \\
$dt$          & 0.001  & 0.005 & time step \\
$\epsilon^{(0)}$         & -0.2  & -0.2 & initial guess \\
\br
\end{tabular}
\end{indented}
\end{table}
%
The parameters used in the runs are summarized in \tref{tab:parameter}. The initial guess for the laser field was $\epsilon^{(0)}(t)=-0.2$ in all calculations. This choice is arbitrary but has the advantage of producing a significant occupation in the target state at the end of the pulse, necessary to get the iteration working. Although the simple choice $\epsilon^{(0)}(t)=0.0$ will work as well in most cases, it represents a minimum of the functional since initial and target state are orthonormal. Therefore the algorithm could get stuck in principle.
The obtained solutions, which are presented in the following chapter, are all far away from the initial guess. This suggests that the solutions do not depend on the initial guess for the laser field. 
%

\section{Results}
\label{sec:results}
%
In this section we apply the algorithms described above to our 1D model for electron transfer. We start in the ground state $| 0 \rangle$ ($t=0$) where the electron is localized in the left well and demand that at the end of the laser pulse ($t=T$) it will be transfered to 1st excited state $|1 \rangle$, which is mainly located in the right well (see \fref{fig:potential_states}). The target operator in this case is a projection operator onto the first excited state: $\widehat{O} = | 1 \rangle \langle 1 |$. Therefore, the success is measured by $| \langle \Psi(T) | 1 \rangle |^2$ which we simply refer to as the ``yield''.
The pulse length is chosen to be $T=400$ ($\approx 9.7$~fs).

%
\subsection{Fluence constraints} \label{sec:fluence}
%
\subsubsection{Fixed fluence.}\label{sec:fluence_2lev}
\begin{figure}[!h]
\centering 
\subfigure[]{
  \label{fig:field_E1}
      \includegraphics*[width=.33\textwidth]{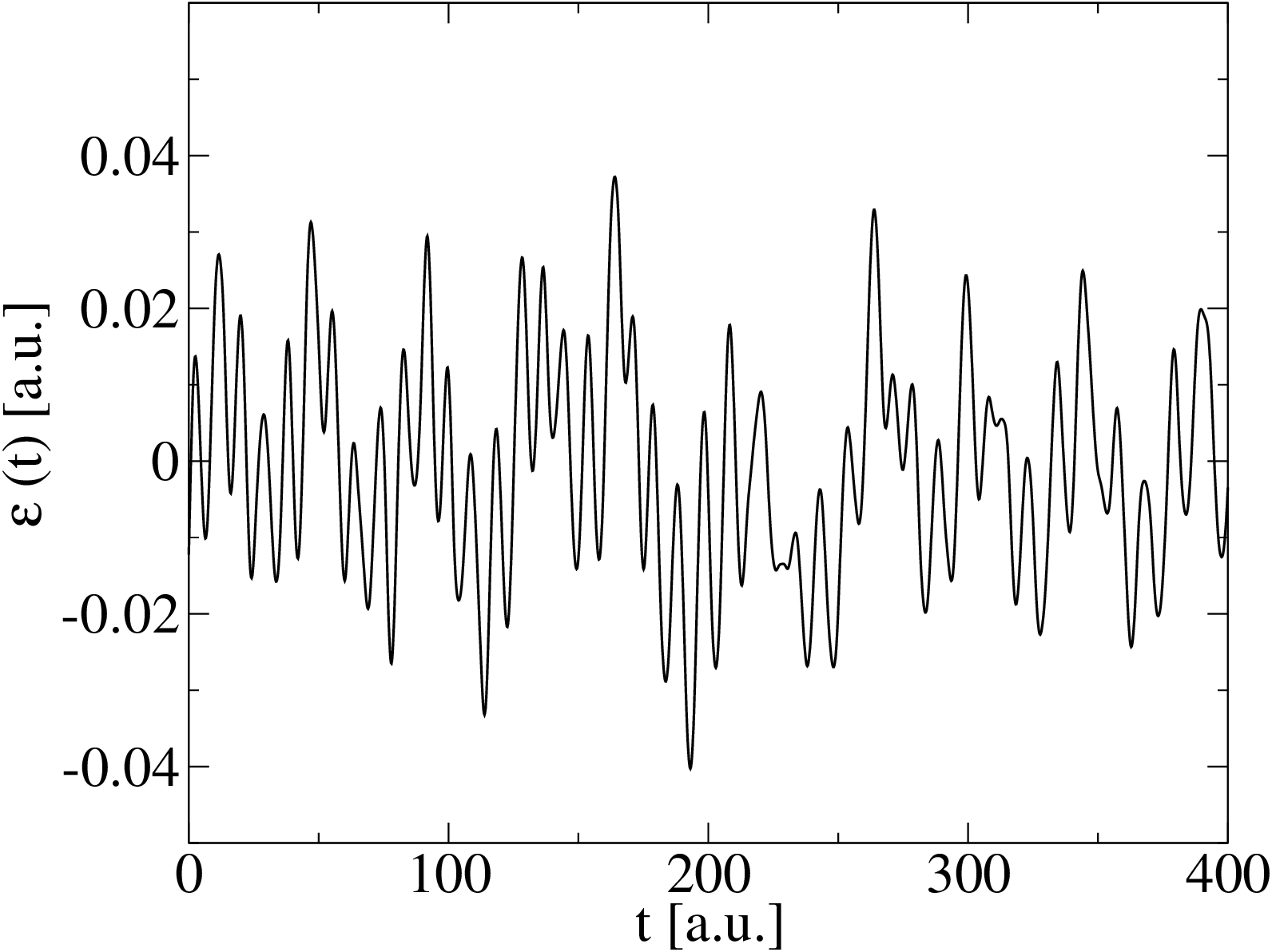}
    }
\subfigure[]{
  \label{fig:field_E2}
  \includegraphics*[width=.33\textwidth]{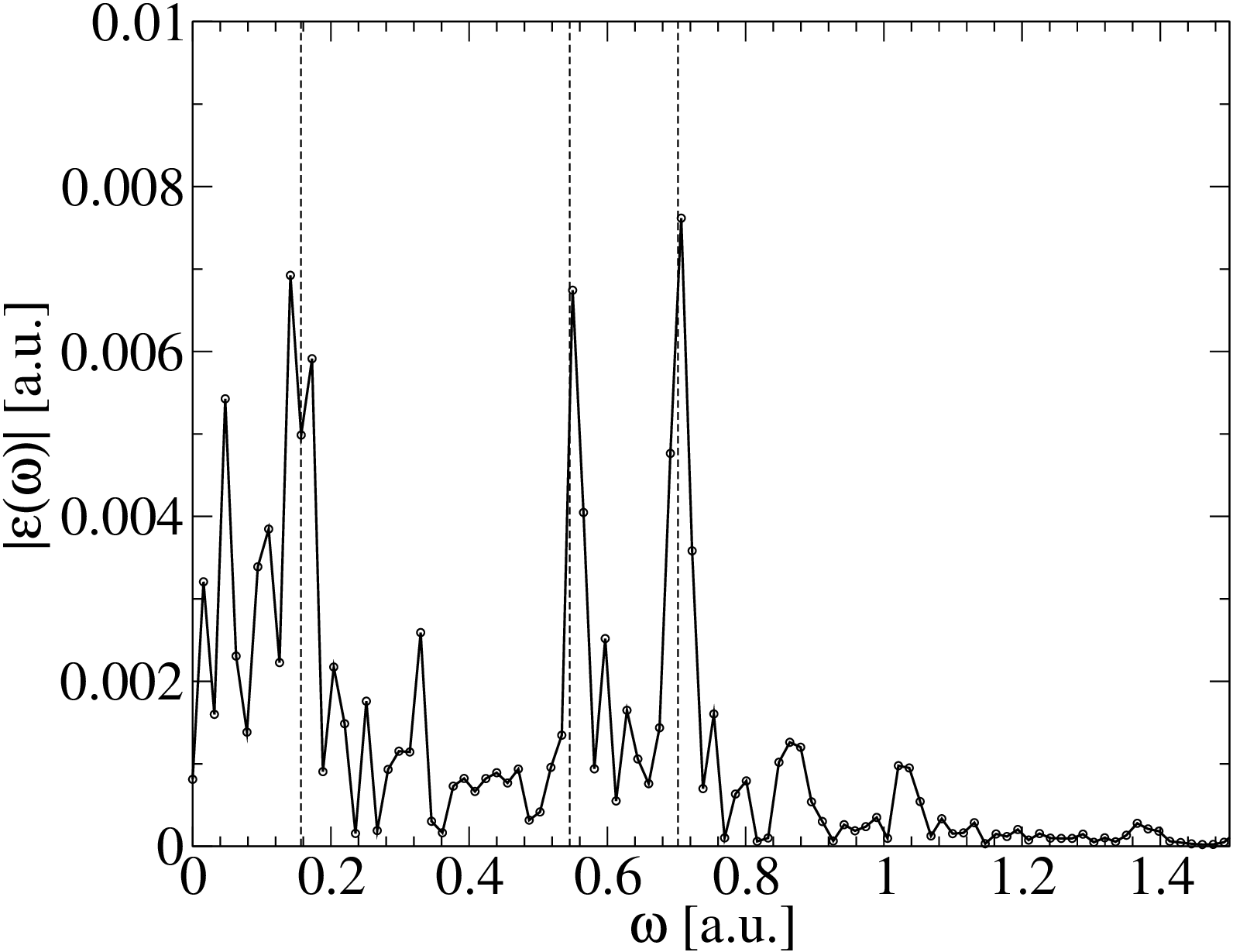}
}
\subfigure[]{
  \label{fig:field_E3}
  \includegraphics*[width=.33\textwidth]{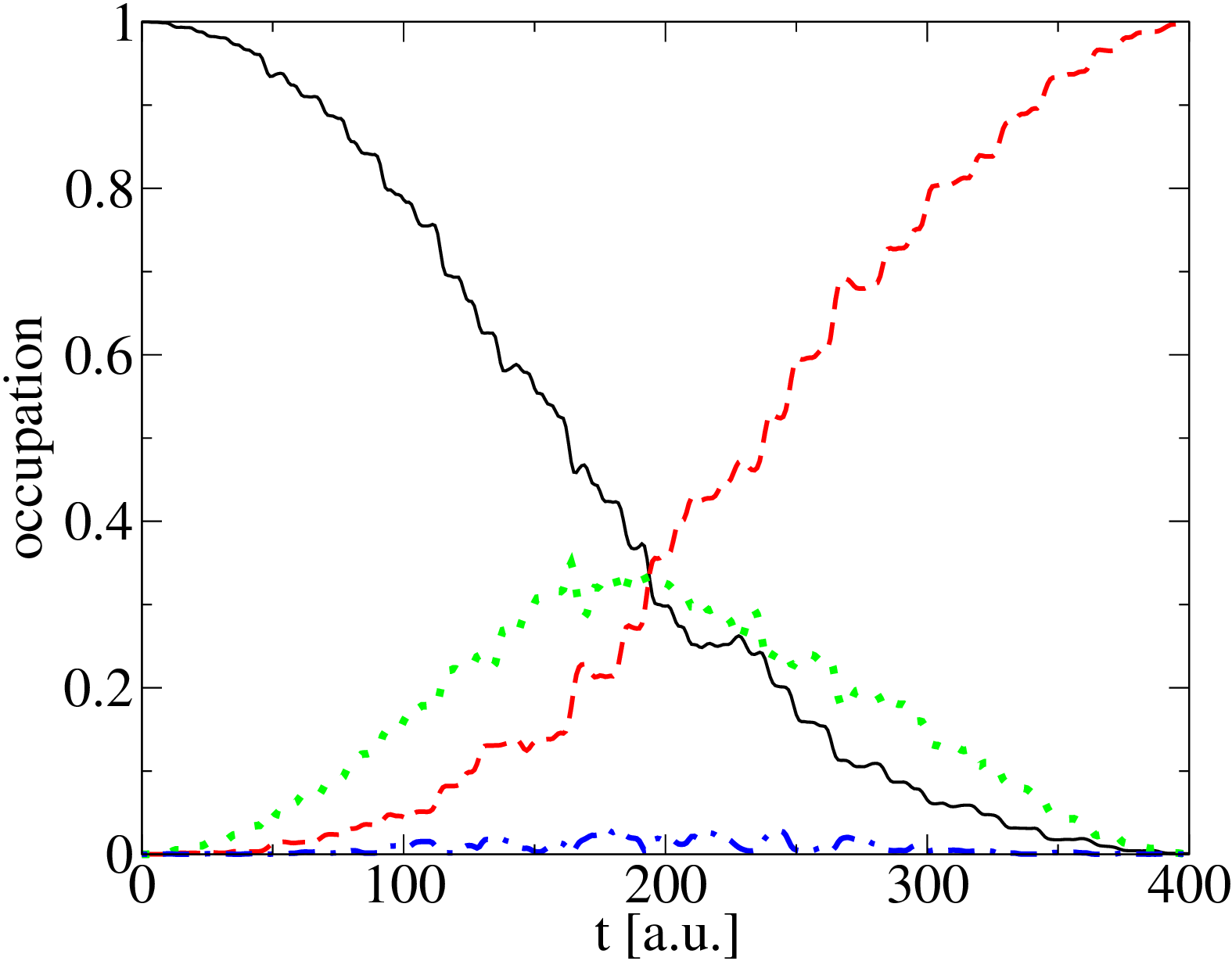}
}
\subfigure[]{
  \label{fig:field_E4}
  \includegraphics*[width=.33\textwidth]{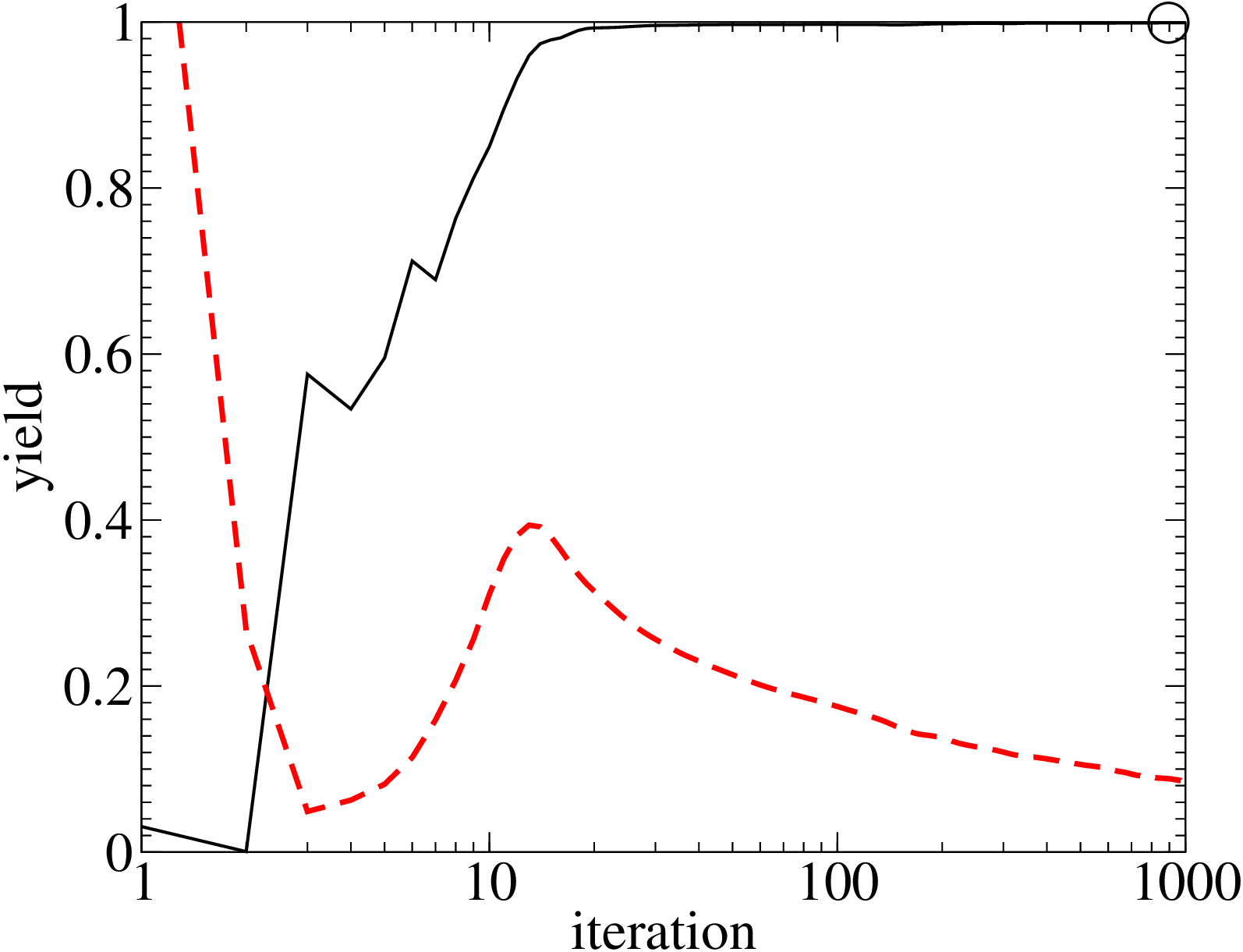} 
}
\caption{We apply the algorithm for $E_0=0.080$. The optimized field is shown in (a), with its spectrum in (b). In (c) we plot the time evolution of the occupation numbers $|\langle \Psi(t) | n \rangle |^2$ ($n=0$ (\full), $n=1$ (\dashed), $n=2$ (\dotted) and $n=3$ (\chain)). The convergence behavior is shown in (d). Also shown in (d) are the values of the Lagrange multiplier $\alpha$ (\dashed) during the iteration which we scaled by a factor of $0.1$. The \opencircle indicates the iteration with the highest yield.}
\label{fig:field_E}
\end{figure}

%
In the following we first apply our algorithm to find an optimal field with the fluence $E_0=0.080$. This is the value obtained by an estimate using the two-level system (see \ref{app_2lev}). After $894$ iterations we obtain a yield of $99.91\%$ which is higher than the yield found by the two-level estimate. The optimal laser field and its spectrum (Fourier transform) are shown in figures \ref{fig:field_E1} and \ref{fig:field_E2}. The spectrum is dominated by three narrow peaks which correspond to the excitation energies $\omega_{01}=0.156$, $\omega_{12}=0.545$ and $\omega_{02}=0.702$. This suggests that the optimized transition process is a mixture of the direct process, i.e. the excitation from $|0\rangle \to |1\rangle$ and an indirect process which uses the second excited state as intermediate state: $|0\rangle \to |2\rangle \to |1\rangle $. Other indirect processes, like $|0\rangle \to |3\rangle \to |1\rangle $, play only a minor role in this case. This interpretation is supported by looking at the evolution of the occupation numbers in time (\fref{fig:field_E3}). First, the laser pulse populates the second excited state (\dotted) and then after half of the pulse duration depopulates it again.  
Looking once more at the spectrum (in \fref{fig:field_E2}) we observe a group of peaks around $\omega_{01}$ 
($\omega\in [0,0.1]$ and $\omega \in [0.2,0.3]$) 
which do not correspond to any excitation energy of the field-free Hamiltonian. However, these frequencies play an important role in the transition process. If we filter out these frequency components, rescale the fluence to $E_0=0.080$, and then propagate this modified laser pulse, we find, at the end of the pulse, the following occupations: ground-state $16\%$, first excited state $38\%$, second excited state: $44\%$, and in all higher levels $6\%$. In particular, the direct transition and the back transfer from the intermediate level $|2 \rangle $ to the target state in the indirect process are less efficient without these extra frequencies. Further analysis of this kind shows that the low-frequency components and especially the zero frequency component (bias) are crucial since they introduce a (slight) shift of the resonance frequencies, visible as a broadening of the $\omega_{01}$ peak in \fref{fig:field_E2}. If, on the other hand, these components are missing the remaining frequencies become slightly off-resonant, resulting in the low efficiency of $44\%$.   

If we filter out everything except the extra peaks we find a target state occupation of $1\%$. Understanding these extra peaks as a third type of transfer process (see \sref{sec:lowfrequency}) suggests that, in this case, a mixing of transition processes seems to be superior in terms of the maximum target yield than a pulse consisting of a single process only, e.g. the direct process. 

The final yield $99.91\%$ is only $0.61\%$ better than the yield coming from the simple monocromatic pulse estimate of the two-level system. This gain has a high price, the optimized pulse is hardly realizable in any experiment. Although the gain improves with shorter pulse lengths (see \ref{app_2lev}), this example demonstrates the typical dilemma between theory and experiment: Calculated pulses often have a far too complicated spectrum to be produced in practice. In \sref{sec:spectral} and \ref{sec:combination} we demonstrate how this dilemma can be resolved.

To conclude the analysis we look at the convergence behavior of the applied scheme (see \fref{fig:field_E4}). We find a fast convergence within the first 20 iterations. After these 20 iterations the improvement of the yield slows down, like it is also found in the rapid monotonic schemes presented in references \cite{ZBR98,MT2003}.

\subsubsection{Energy versus yield}
%

\begin{figure}[!h]
\centering
\includegraphics*[width=.43\textwidth]{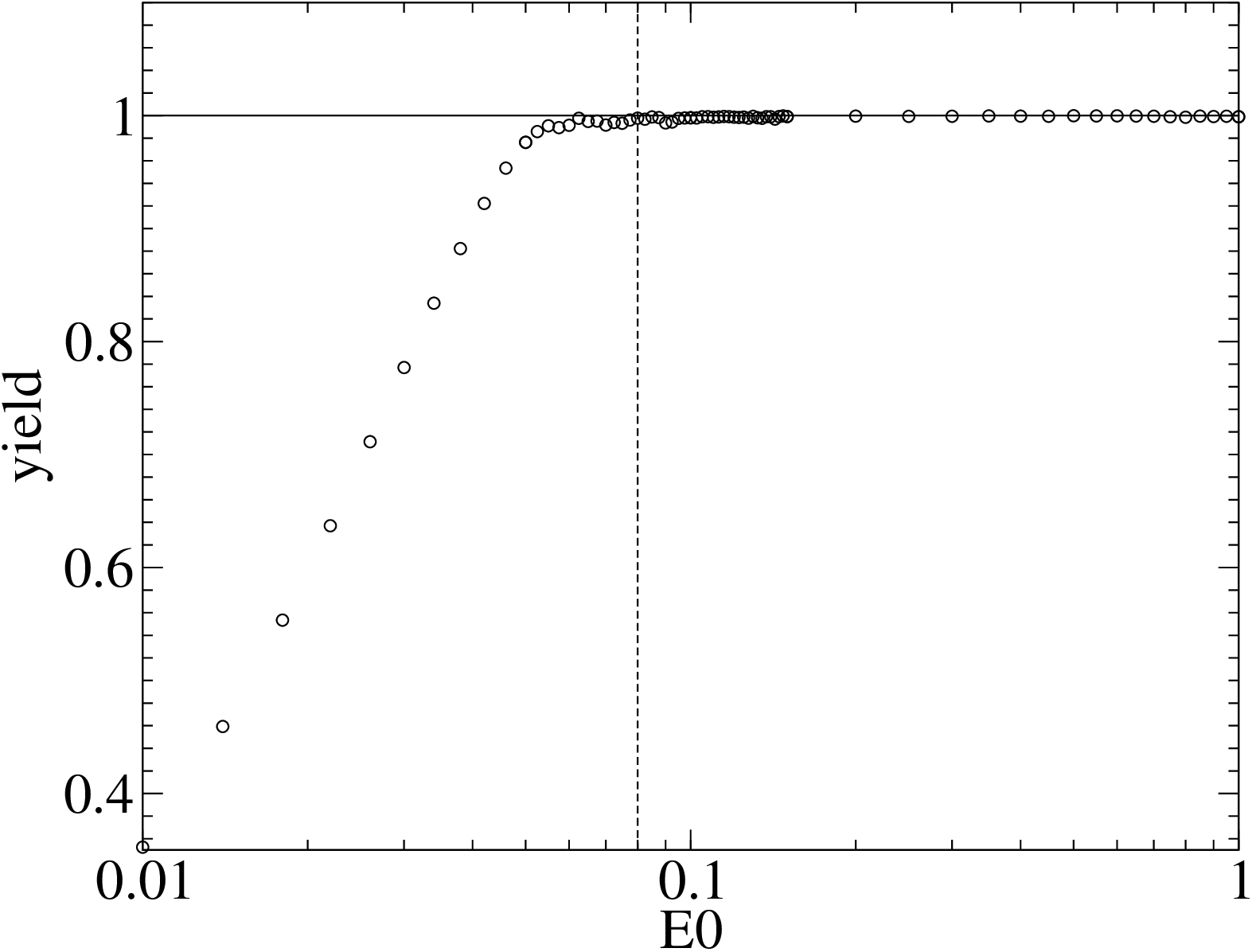}       
\caption{The graph shows the yield for different values of $E_0$ ranging from $0.010$ to $1.000$. Each point corresponds to a single optimization. For these runs we have used a smaller grid ($x_{max}=-x_{min}=20.0$, $dx=0.15625$) and a larger time-step ($dt=0.005$). The vertical line (\dashed) corresponds to $E_0=0.080$. Beyond this line only yields higher than $0.99$ are found.} 
\label{fig:scan_En}
\end{figure}

We apply our method to scan through a range of values for $E_0$ from $0.010 \ldots 1.000$. 
The scan, displayed in \fref{fig:scan_En}, shows that there seems to be a critical value $\tilde{E}_0$ which is necessary to get very high occupations ($|\langle \Psi(T)|1 \rangle |^2 > 0.99$ of the target state. For values $E_0>\tilde{E}_0$ the algorithm always finds a laser field that produces yields above $99\%$.

For long pulse durations (as it is the case here) we can give a rough estimate of this critical value  $\tilde{E}_0$  with the help of the two-level system:
\begin{eqnarray}
\tilde{E}_0 &\approx & A^2 \frac{T}{2} = \frac{\pi^2}{2 \mu_{01}^2 T }\\
\end{eqnarray}
with $A =  \pi / \left(\mu_{01} T\right)$ (see \ref{app_2lev}).
%

\begin{figure}[!h]
\centering 
\subfigure[]{
  \label{fig:fields_Escan1}
      \includegraphics*[width=.4\textwidth]{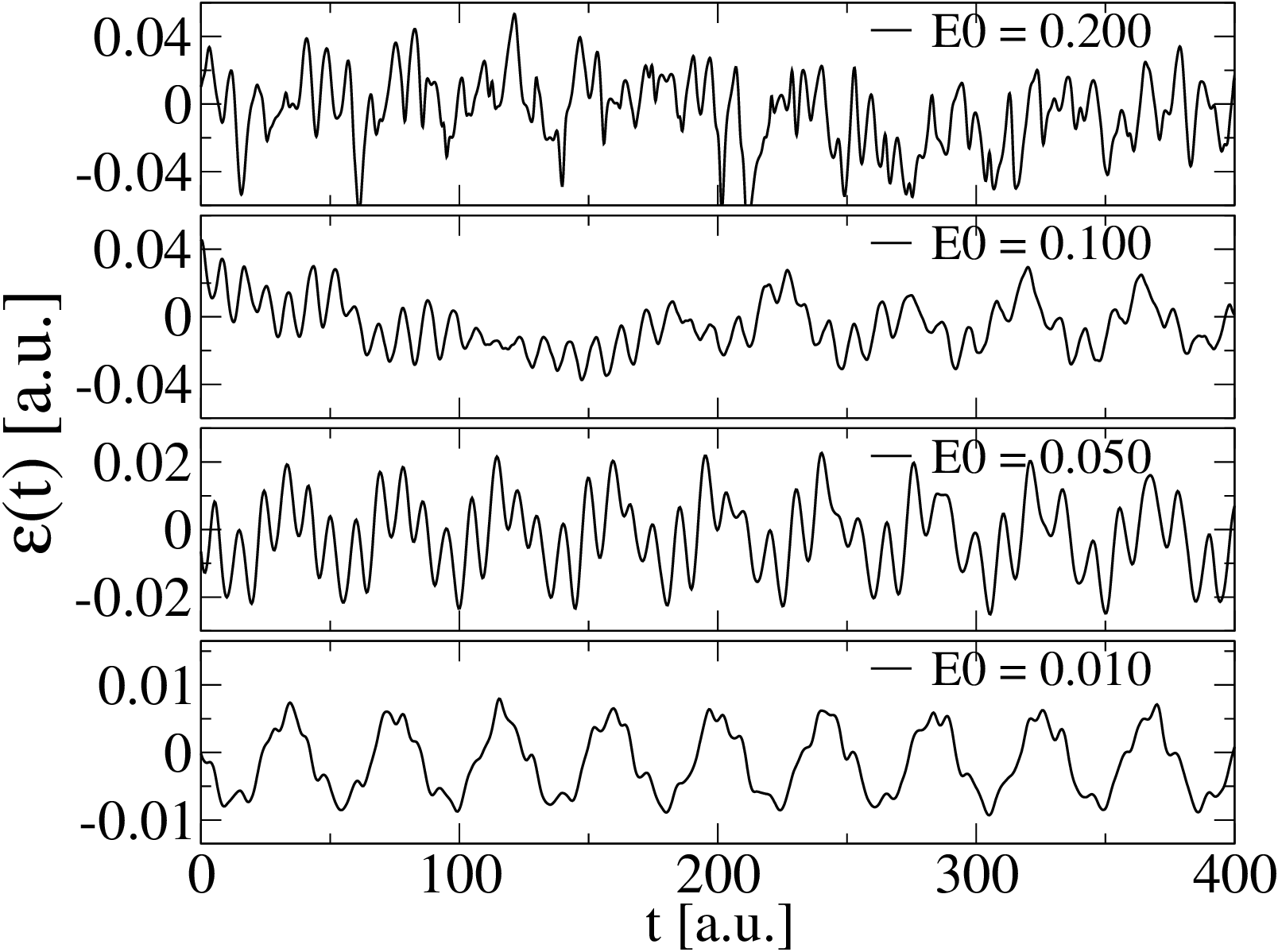}
      \hspace{.3in}
    }
\subfigure[]{
  \label{fig:fields_Escan2}
  \includegraphics*[width=.4\textwidth]{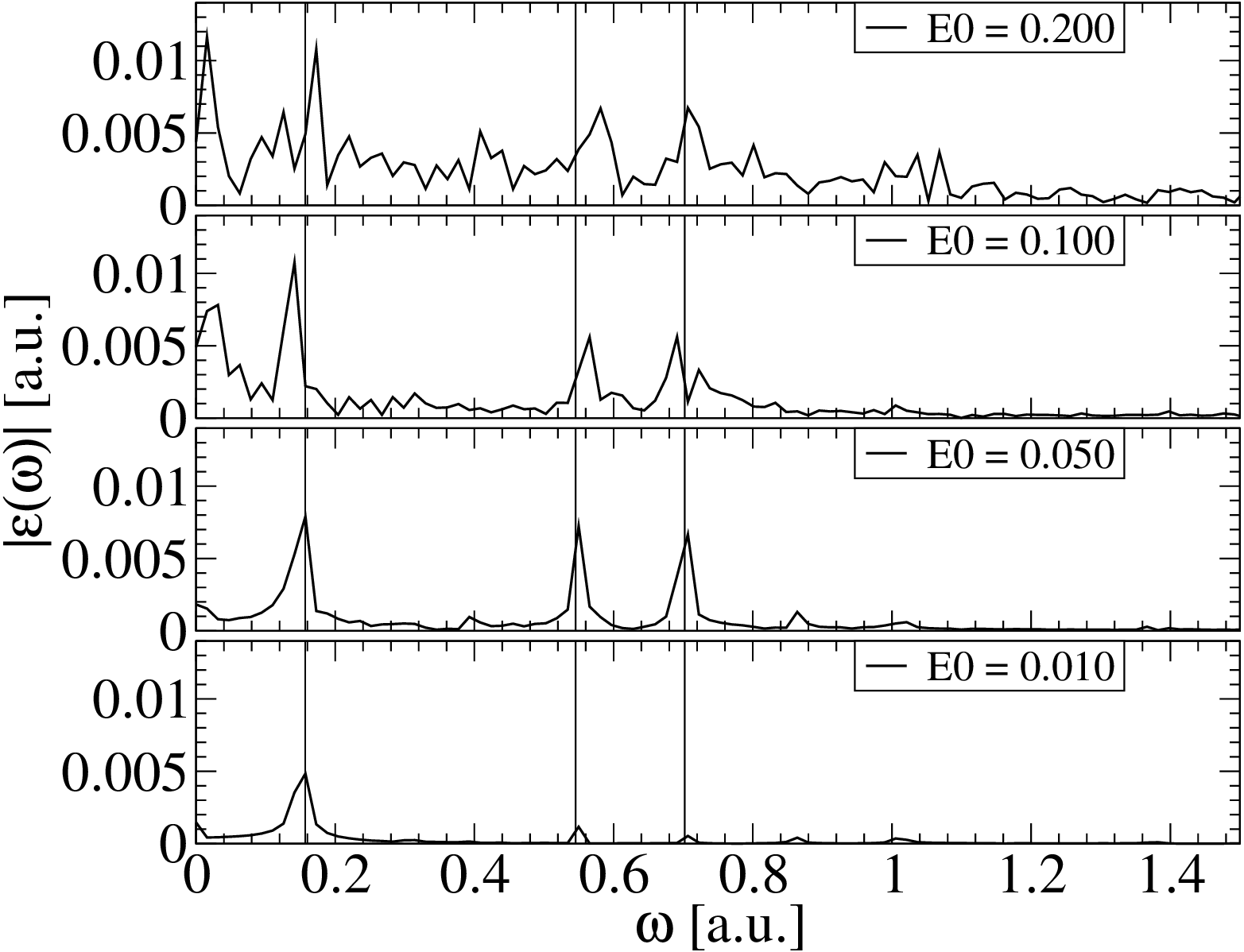}
\hspace{.3in}
}
\caption{In (a), we plot the optimized laser fields for different values of $E_0={ 0.010, 0.050, 0.100, 0.200 }$. Graph (b) shows the corresponding spectra. The resonance frequencies for the transitions $|0 \rangle \to |1 \rangle$, $|1 \rangle \to |2 \rangle$ and  $|0 \rangle \to |2 \rangle$  are indicated by vertical lines. In the two upper plots of graph (b) the peaks are Stark-shifted. Note, that these spectra also contain a large low frequency part.}
\label{fig:fields_Escan}
\end{figure}

If we take a closer look at some of the optimized fields (see \fref{fig:fields_Escan1}) for the values  $E_0=0.010,0.050,0.100,0.200$, we see that the spectra (see \fref{fig:fields_Escan2}) of these pulses get more complicated as the assigned fluence increases. In the lower two panels ($E_0=0.010,0.050$) we find peaks at the exact resonance frequencies. The optimized fields result in occupations of $35.24\%$ and $97.63\%$. While the pulses shown in the two upper panels ($E_0=0.100,0.200$) produce yields of $99.81\%$ and $99.95\%$. The peaks corresponding to the direct $|0\rangle \to |1 \rangle$  and indirect process $|0\rangle \to |2 \rangle \to |1\rangle $ are ``Stark'' shifted. For the stronger pulses, we also find an increasing low-frequency part. The spectrum in the top panel ($E_0=0.200$) is difficult to analyze. However, one can see that more and more processes are taking part in the transition, i.e. peaks occur near the other resonance frequencies.

%
\subsection{Spectral constraints} \label{sec:spectral}
%
%
In the following we present the results of the algorithm with spectral constraints and penalty factor for two examples of the filter function. These examples are motivated by the findings of the previous chapter, namely that the transfer of the particle occurred via a mixture of a direct transition and indirect transitions. We want to find a laser pulse that produces a high yield and only contains spectral components centered around the resonance frequency $\omega_{01}$. We know that such a pulse exists, since it appeared in the second iteration when looking for a pulse with $E_0=0.080$ (see \sref{sec:fluence_2lev}). 
In the second example we optimize a laser that is not allowed to contain the excitation frequency $\omega_{01}$ of the direct process.

%
\subsubsection{Direct transition} \label{sec:direct}
%
Using spectral constraints in the optimization scheme allows us to explicitly select the direct transition, i.e. we search for a pulse whose main frequency component is the excitation energy $\omega_{01}$. This is done by applying a Gaussian shaped frequency filter $f(\omega)$, according to equation \eref{eq:filter1}, centered around $\omega_0 = \omega_{01}$ and with $\gamma=500$.  

\begin{figure}[!h]
\centering 
\subfigure[]{
  \label{fig:field_F1}
      \includegraphics*[width=.33\textwidth]{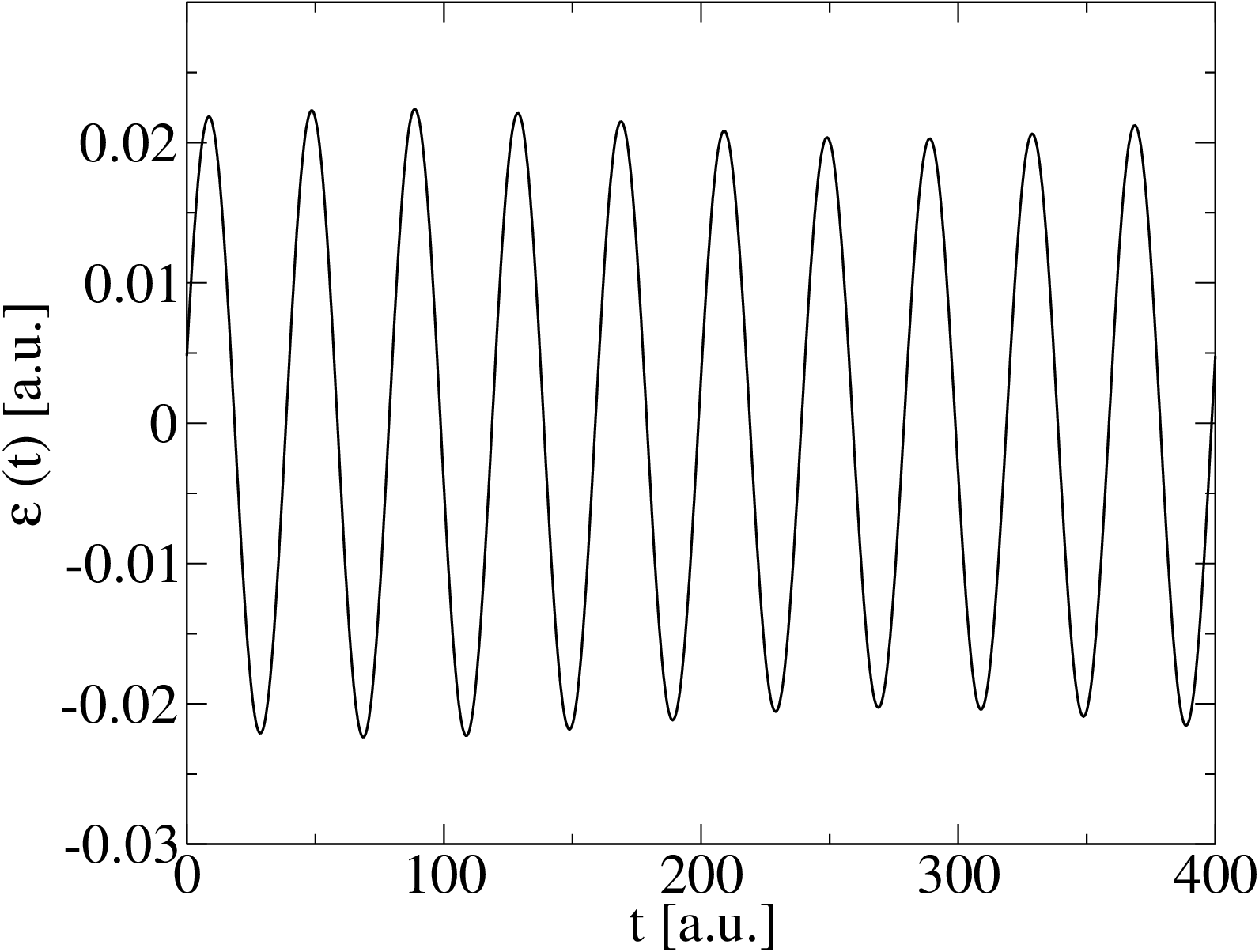}
    }
\subfigure[]{
  \label{fig:field_F2}
  \includegraphics*[width=.33\textwidth]{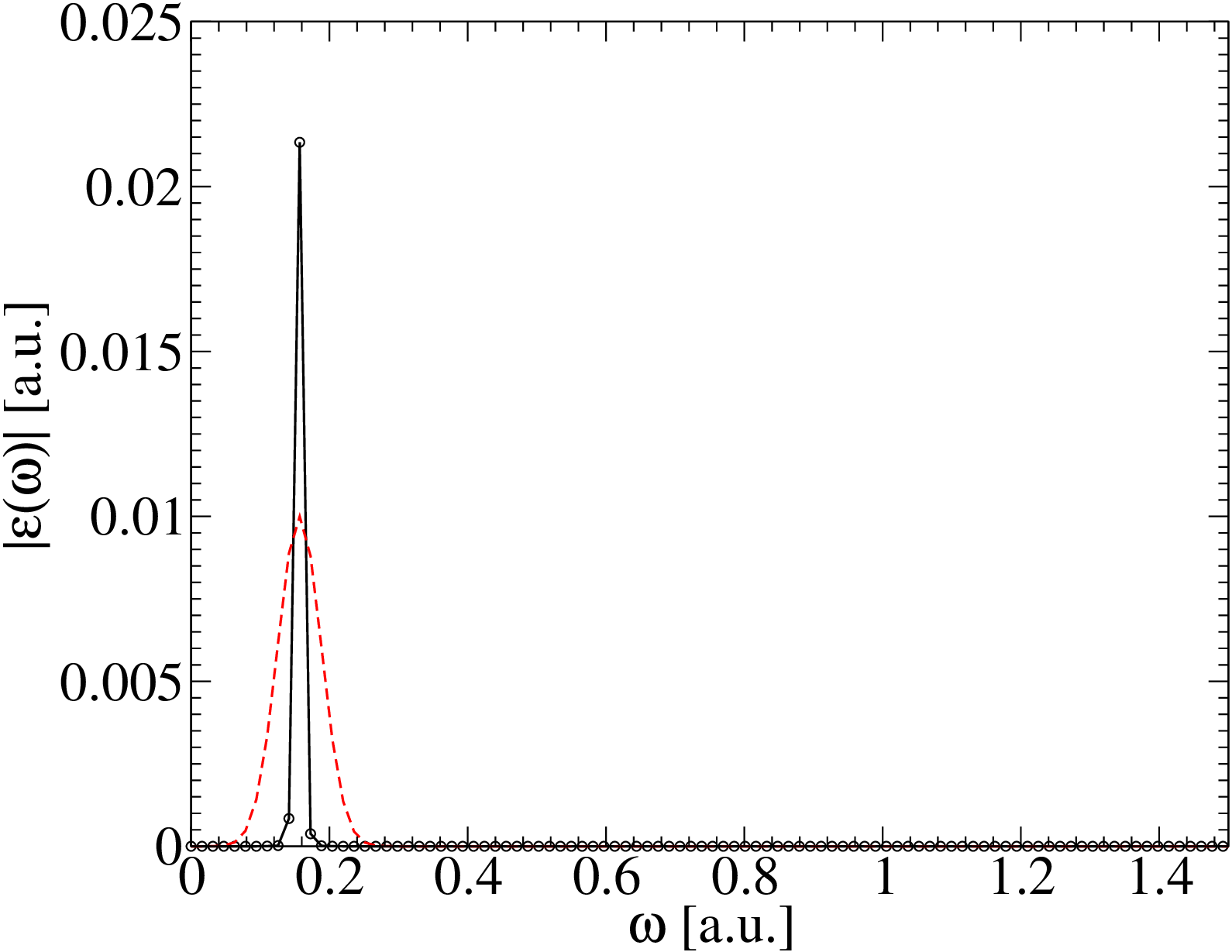}
}
\subfigure[]{
  \label{fig:field_F3}
  \includegraphics*[width=.33\textwidth]{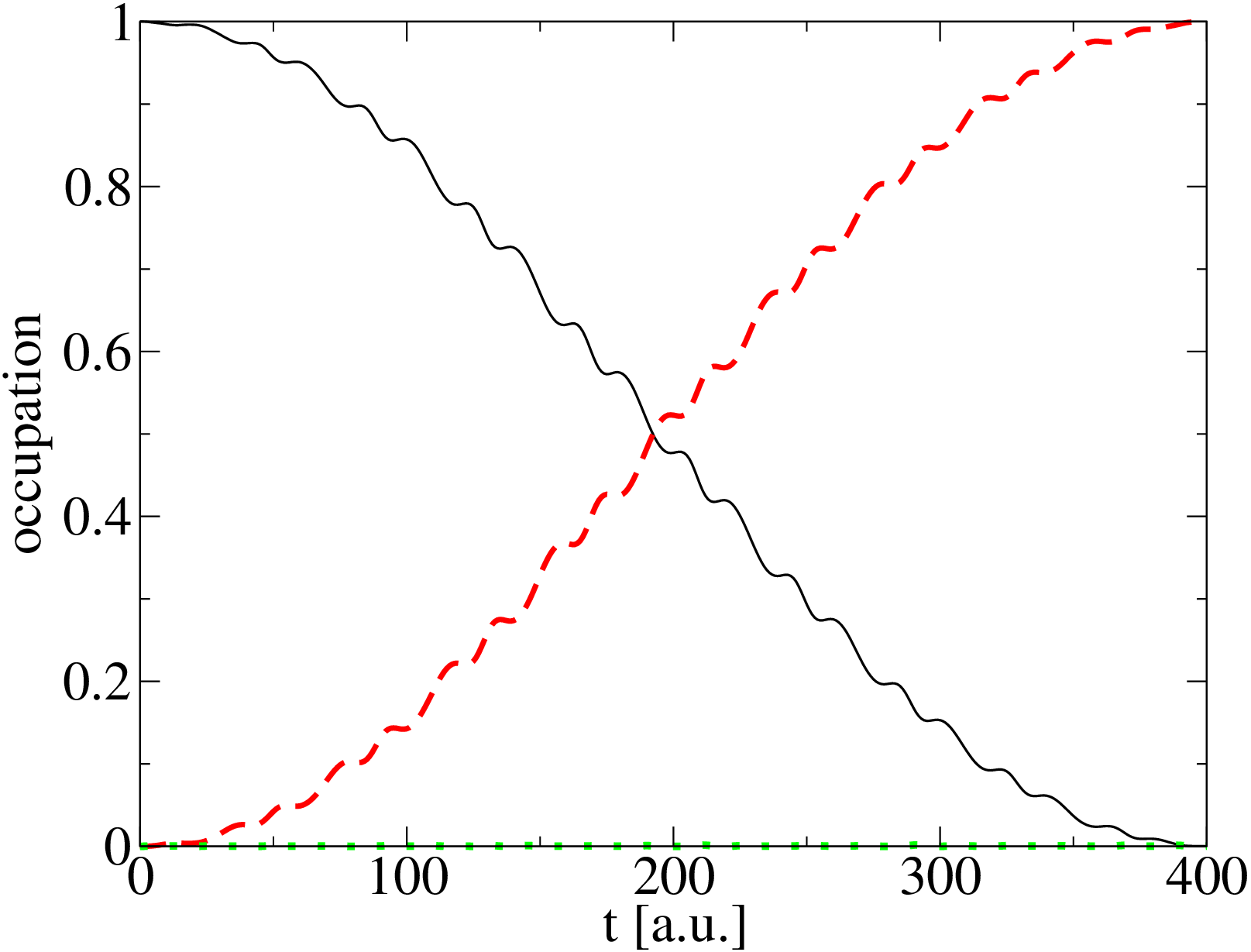}
}
\subfigure[]{
  \label{fig:field_F4}
  \includegraphics*[width=.33\textwidth]{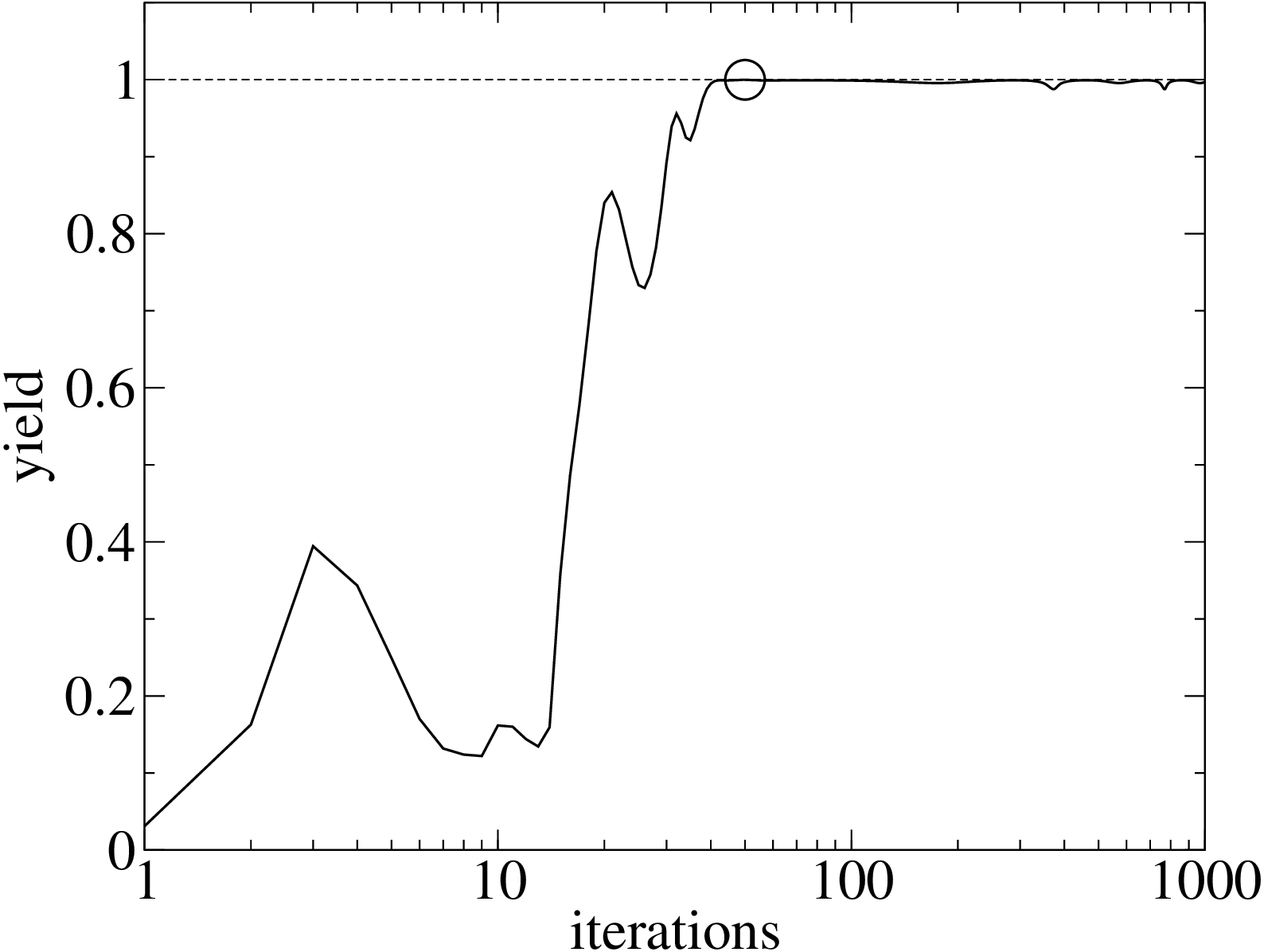}
}

\caption{We apply the optimization algorithm but allowing only a Gaussian frequency distribution around $\omega_{01}=0.1568$. The resulting field is displayed in graph (a). The filter function $f(\omega)$ (\dashed) which is scaled by $0.01$ is shown together with the spectrum in (b). The time dependent occupation numbers (c) confirm that only the ground state (\full) and the first excited state (\dashed) take part in the process. The second excited state population (\dotted) is hardly visible. The convergence is shown in graph (d). The \opencircle indicates the iteration with the highest yield.}
\label{fig:field_F}
\end{figure}

After $50$ iterations the algorithm finds a laser pulse which results in a yield of $99.97\%$ . We set the penalty factor $\alpha=0.05$ and obtain a value of $E_0 = 0.090$ which is slightly higher than the estimate from the two-level model but also more effective.
The slight envelope on the field, shown in \fref{fig:field_F1}, stems from the finite width of the Gaussian (see \fref{fig:field_F2}). Frequency components near $\omega_{01}$ are still allowed in the pulse and result in a beat pattern. The time dependent occupation numbers confirm that the higher states are not occupied during the transition (see \fref{fig:field_F3}). The convergence, shown in \fref{fig:field_F3} is rather smooth.
Note, that if we desire a sinusoidal field with a constant envelope we have to reduce the width of the Gaussian to allow only one single component in the spectrum (a Kronecker delta). Using such a filter we obtain a yield of $99.79\%$ and $E_0=0.085$. The field oscillates with the amplitude $A=0.0207$ which is slightly higher than the amplitude derived from the two-level system (see \ref{app_2lev}).

%
%
\subsubsection{Forbidden direct transition} \label{sec:forbid_direct}
%
By choosing the complement of the filter function from the previous example, i.e., by allowing every frequency component except $\omega_{01}$, we can optimize a field which also produces a very high yield. The filter function is given by equation \eref{eq:filter2} with $\omega_{0} = \omega_{01}$ and $\gamma=500$.

\begin{figure}[!h]
\centering 
\subfigure[]{
  \label{fig:field_FC1}
      \includegraphics*[width=.33\textwidth]{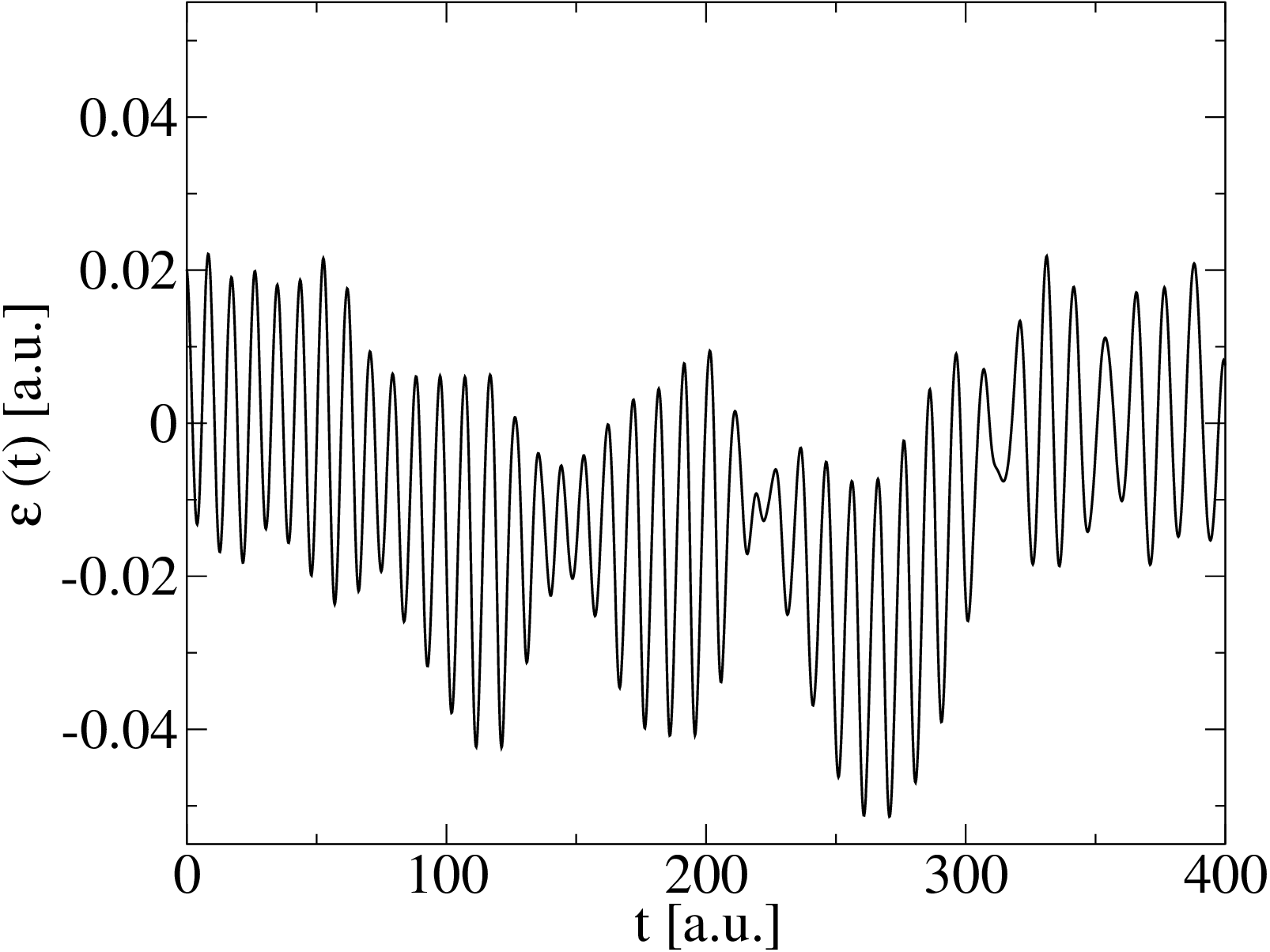}
    }
\subfigure[]{
  \label{fig:field_FC2}
  \includegraphics*[width=.33\textwidth]{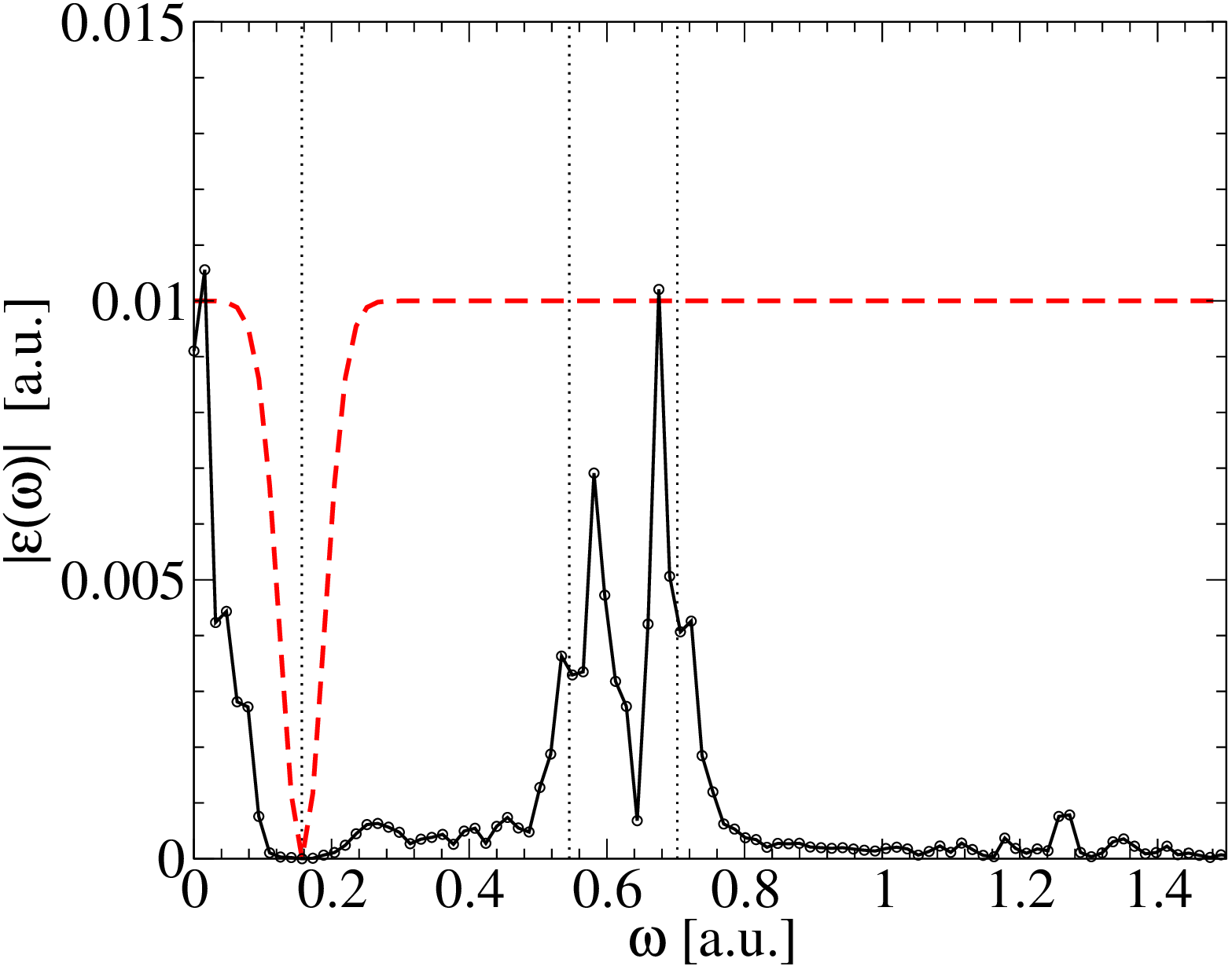}
}
\subfigure[]{
  \label{fig:field_FC3}
  \includegraphics*[width=.33\textwidth]{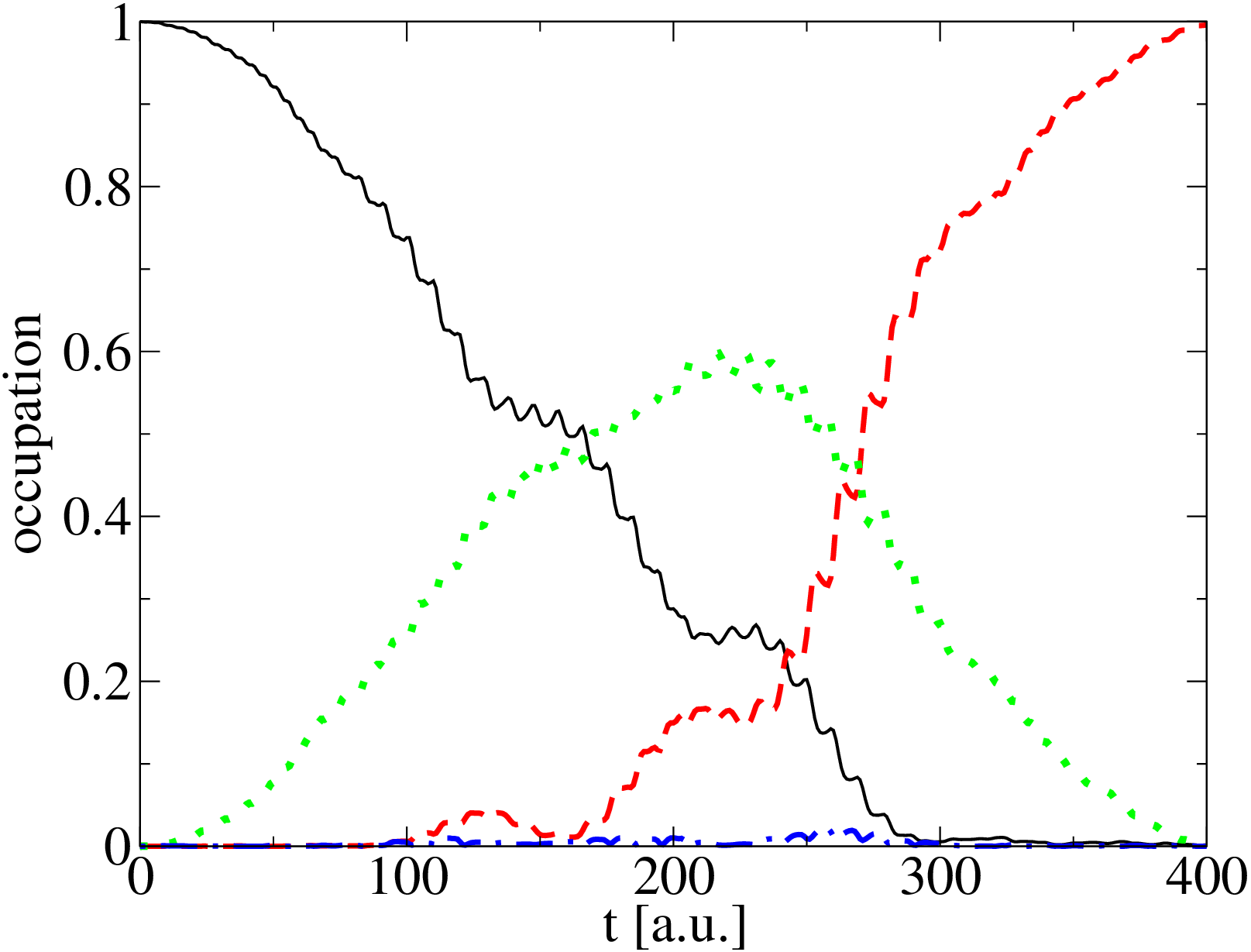}
}
\subfigure[]{
  \label{fig:field_FC4}
  \includegraphics*[width=.33\textwidth]{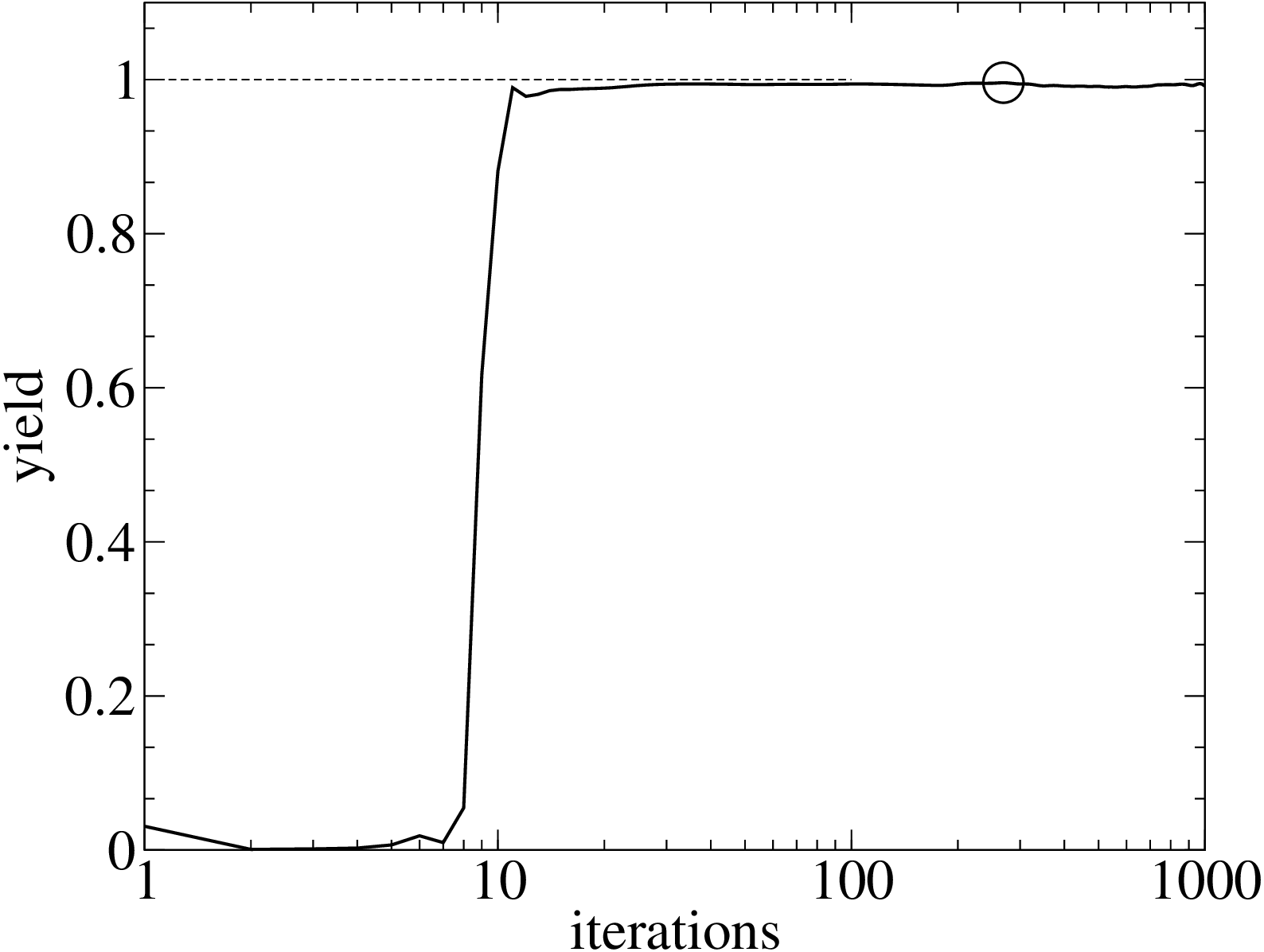}
}

\caption{We prohibit the direct transition from $|0 \rangle \to |1 \rangle $ by using the complement of the previous filter function $\tilde{f}(\omega) = 1 -f(\omega)$. The optimized field is shown in graph (a). The filter function $\tilde{f}(\omega)$ (\dashed), scaled by $0.01$, is plotted together with the spectrum in (b). The time-dependent occupation numbers (c) confirm that now the second excited state (\dotted) plays a major role in the transition (ground state (\full), first-excited state (\dashed) and third excited state (\chain)).  The convergence is shown in graph (d). The \opencircle indicates the iteration with largest occupation of the target-state.}
\label{fig:field_FC}
\end{figure}

The optimization procedure (with a penalty factor $\alpha=2.5$) results in a target state occupation of $99.60\%$ after $269$ iterations. The optimized laser field is presented in \fref{fig:field_FC1}, it integrates to a fluence of $E_0=0.130$. Its spectrum, shown in \fref{fig:field_FC2}, consists of two major components: $\omega_a=0.581$ and $\omega_b=0.676$ which correspond to the Stark-shifted excitation energies $\omega_{12}$ and $\omega_{02}$, i.e. the optimization takes care of the frequency shifts introduced by the large bias (zero-frequency component) of the field. That the transition occurs via the indirect process is confirmed by looking at the time-dependent occupation numbers, shown in \fref{fig:field_FC3}. First, the field starts populating the second excited state and then transfers the population to the target state. Other indirect processes, e.g. $|0\rangle \to |3\rangle \to  |1\rangle$ or $|0\rangle \to |2\rangle \to  |3\rangle \to |1\rangle $, 
play only a minor role: The occupation of the third excited state stays below $2.5\%$ and the frequency components correponding to these processes are very small.
%
\subsection{Combination of Spectral and fluence constraints}
\label{sec:combination}
%
The next examples demonstrate that even more restrictions are possible and we can still obtain very good yields. We combine the spectral restriction with the fluence constraint and continue the above examples by  selecting among the indirect processes. Only two frequencies are allowed in the laser pulse and in addition we fix the fluence.
In the last example we show that it is not even necessary to have resonance frequencies inside the laser pulse to reach very high occupations of the target state. 

%
\subsubsection{Selective transfer via intermediate state $| 2 \rangle$} 
\label{sec:indirect021}
\begin{figure}[!h]
\centering 
\subfigure[]{
  \label{fig:field_FC021_1}
      \includegraphics*[width=.33\textwidth]{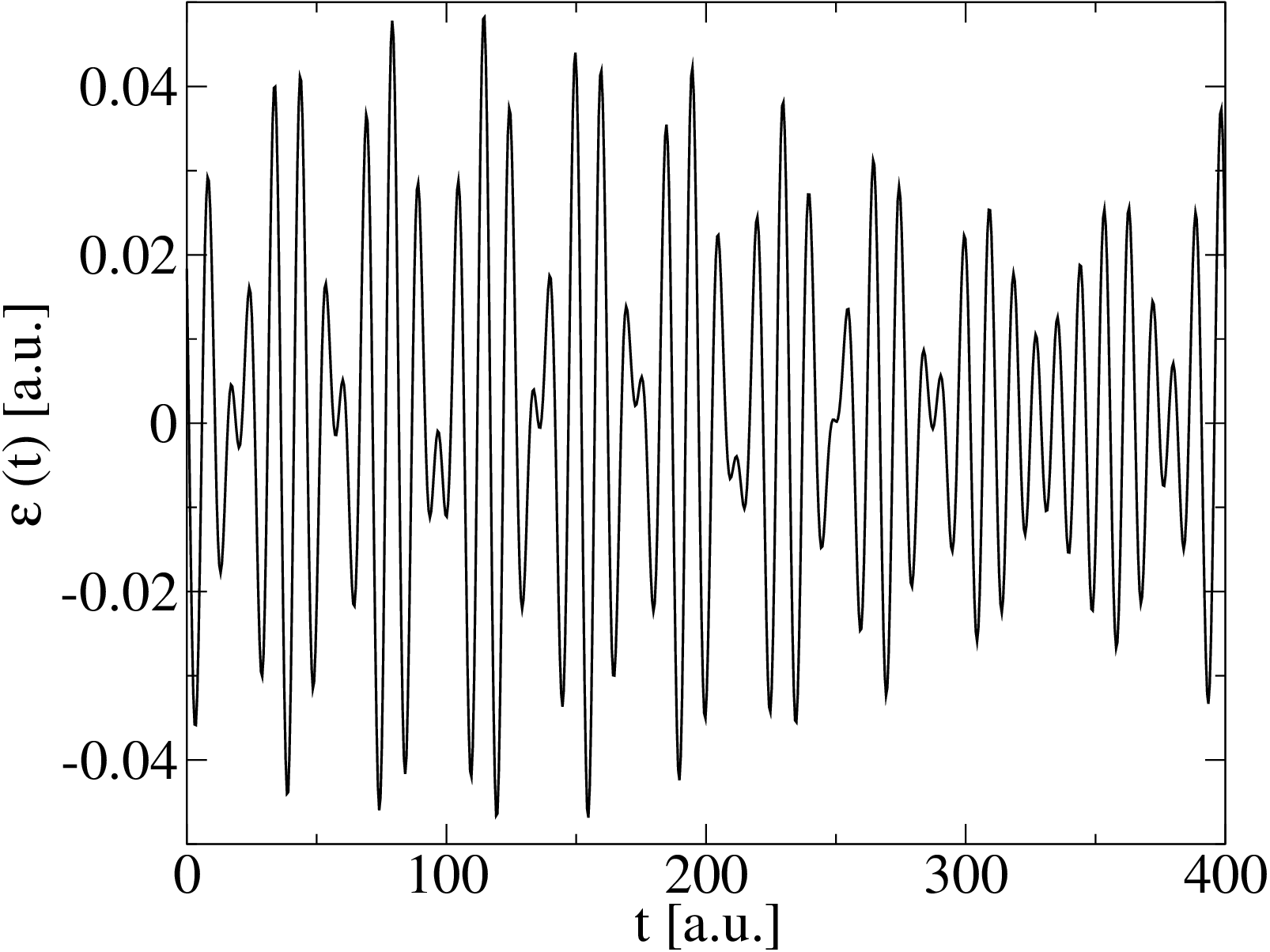}
    }
\subfigure[]{
  \label{fig:field_FC021_2}
  \includegraphics*[width=.33\textwidth]{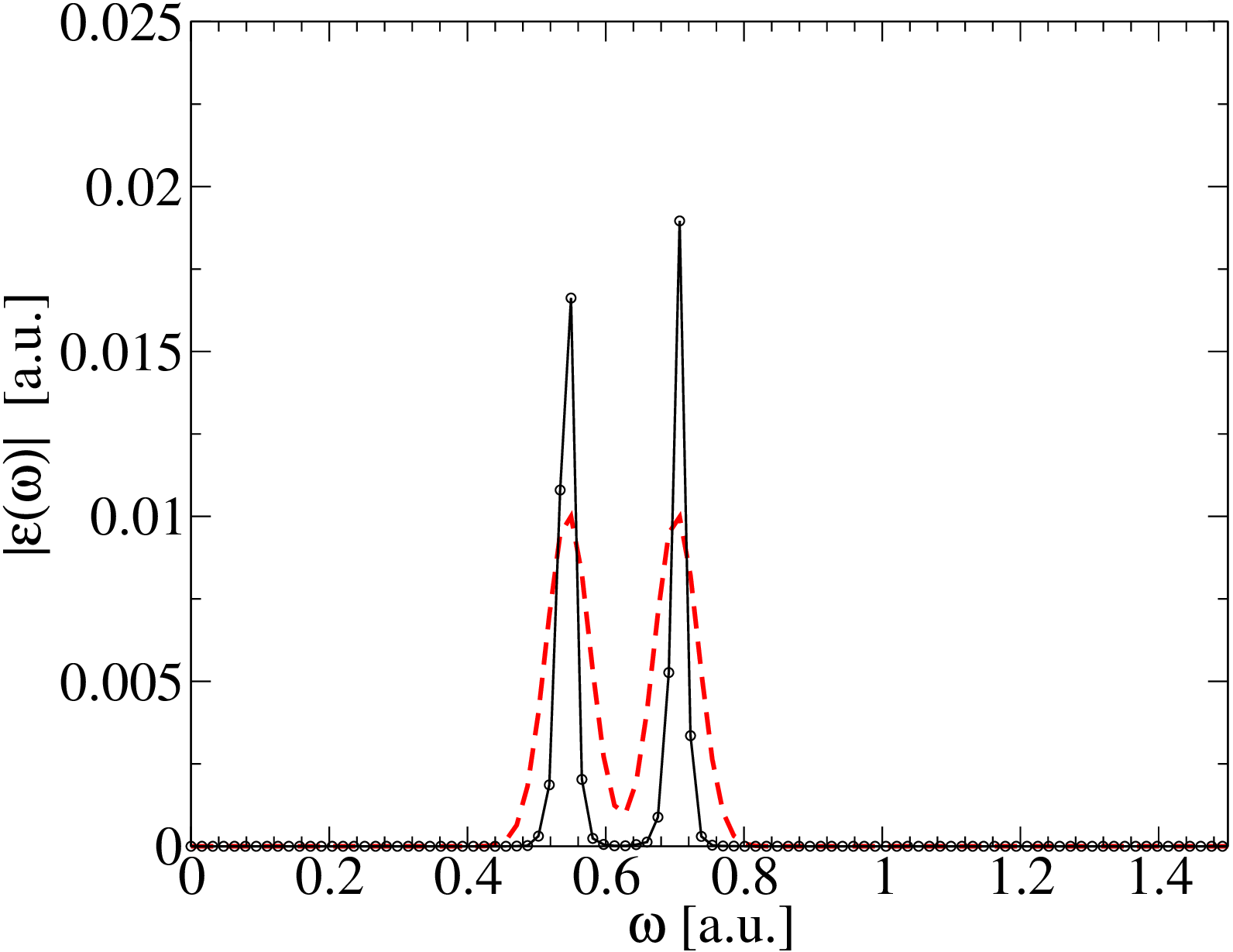}
}
\subfigure[]{
  \label{fig:field_FC021_3}
  \includegraphics*[width=.33\textwidth]{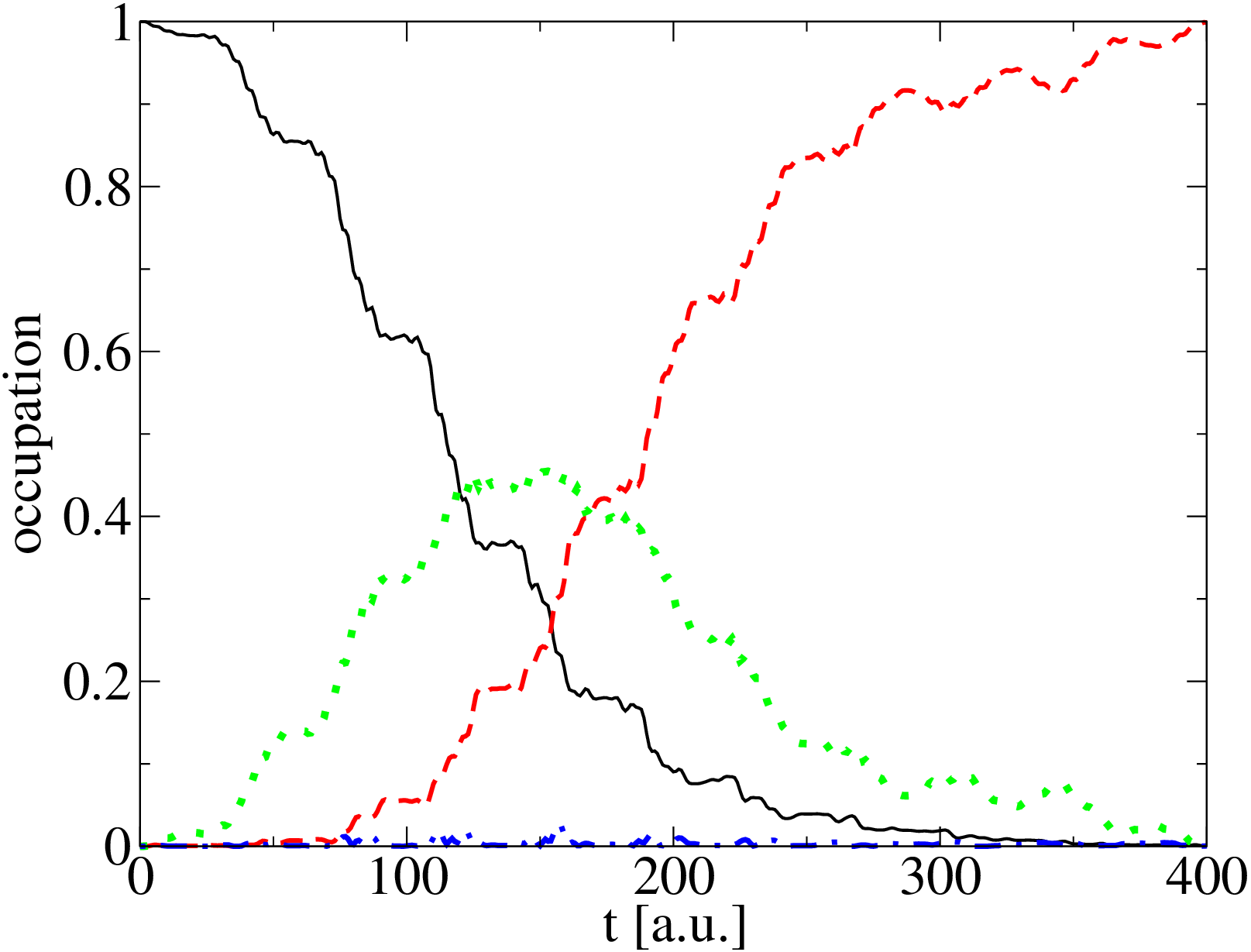}
}
\subfigure[]{
  \label{fig:field_FC021_4}
  \includegraphics*[width=.33\textwidth]{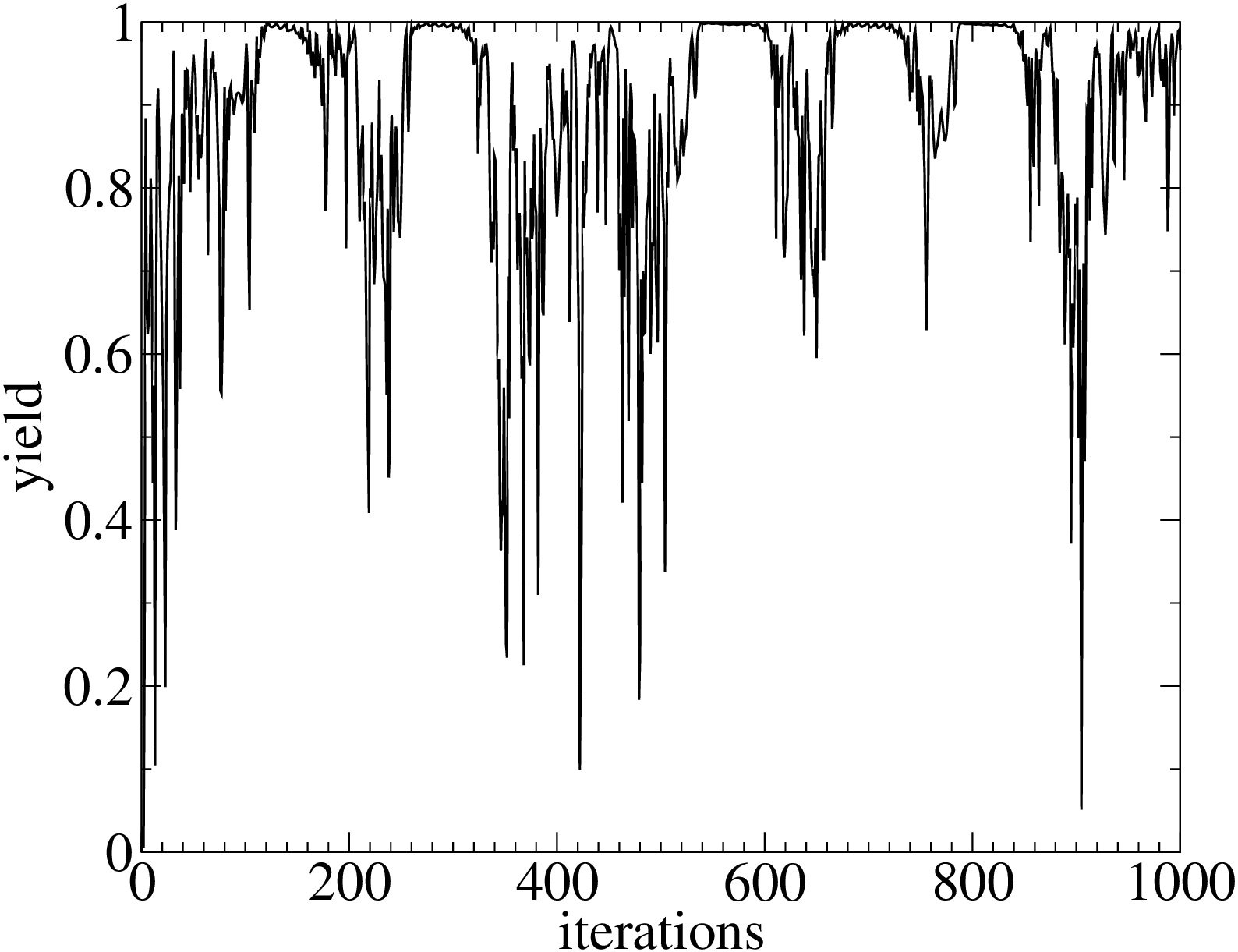}
}

\caption{Here we use a double gaussian window ($\gamma=500$), allowing only frequencies around $\omega_{02}$ and $\omega_{21}$ and in addition demand $E_0=0.160$. The optimized field is shown in graph (a). The filter function (\dashed) $f(\omega)$, scaled by $0.01$, is plotted together with the spectrum in (b). The time-dependent occupation numbers  (ground state (\full), first-excited state (\dashed) and third excited state (\chain)) in (c) confirm that the transition occurs via the second excited state (\dotted). Due to the stronger constraints the convergence behaviour becomes oscillatory, shown in graph (d). }
\label{fig:field_FC021}
\end{figure}

%
In the former examples we found that the indirect process $|0\rangle \to |2\rangle \to |1\rangle$ plays a major role in the excitation process. Since it appeared always together with other processes, e.g. in \sref{sec:fluence_2lev} together with the direct process or in \sref{sec:forbid_direct} together with other indirect processes, we try to find a laser field with only the two excitation energies $\omega_{02}$ and $\omega_{12}$ ($\gamma=500$) and in addition require $E_0=0.160$. For these high requirements we have to pay a price which is the irregular behaviour of the yield during the iteration, shown in \fref{fig:field_FC021_4}. After $540$ iterations we find a yield of $99.90\%$. 
The restriction of the laser frequencies results exactly in the desired transition process, which is confirmed by the time-dependent occupation numbers, shown in \fref{fig:field_FC021_3}. 

\subsubsection{Selective transfer via intermediate state $| 3 \rangle$}
%
The process $|0\rangle \to |3\rangle \to |1\rangle$ using the third excited state as intermediate state played only a minor role in the examples considered so far. Here, we try to optimize the laser pulse so that the transition is only performed via this process. In addition we require $E_0=0.320$. 
Again, we use a double Gaussian filter, one Gaussian centered at $\omega_{13}$, the other one at $\omega_{03}$ and choose the width parameter $\gamma=500$.
 
\begin{figure}[!h]
\centering 
\subfigure[]{
  \label{fig:field_FC031_1}
      \includegraphics*[width=.33\textwidth]{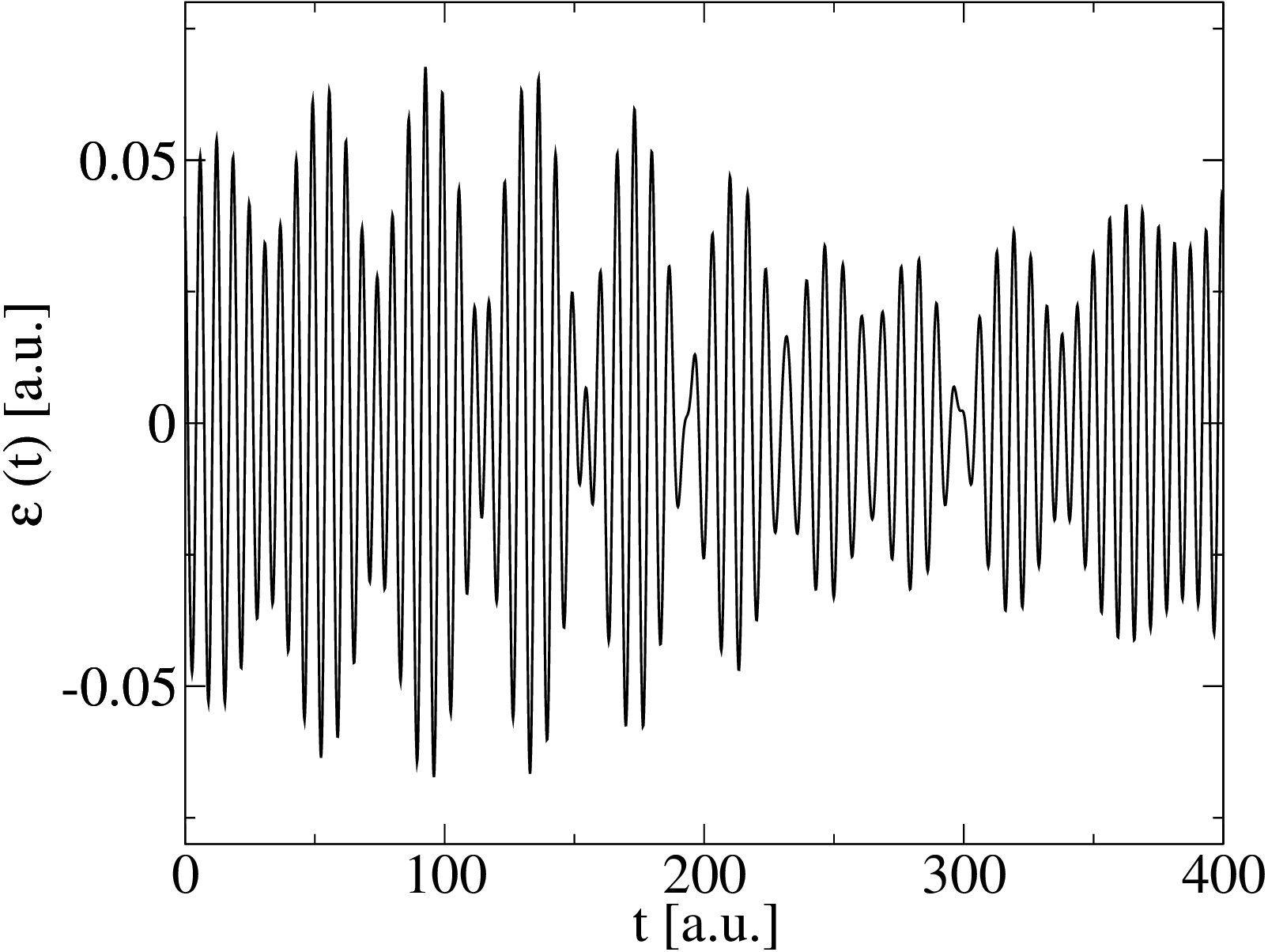}
    }
\subfigure[]{
  \label{fig:field_FC031_2}
  \includegraphics*[width=.33\textwidth]{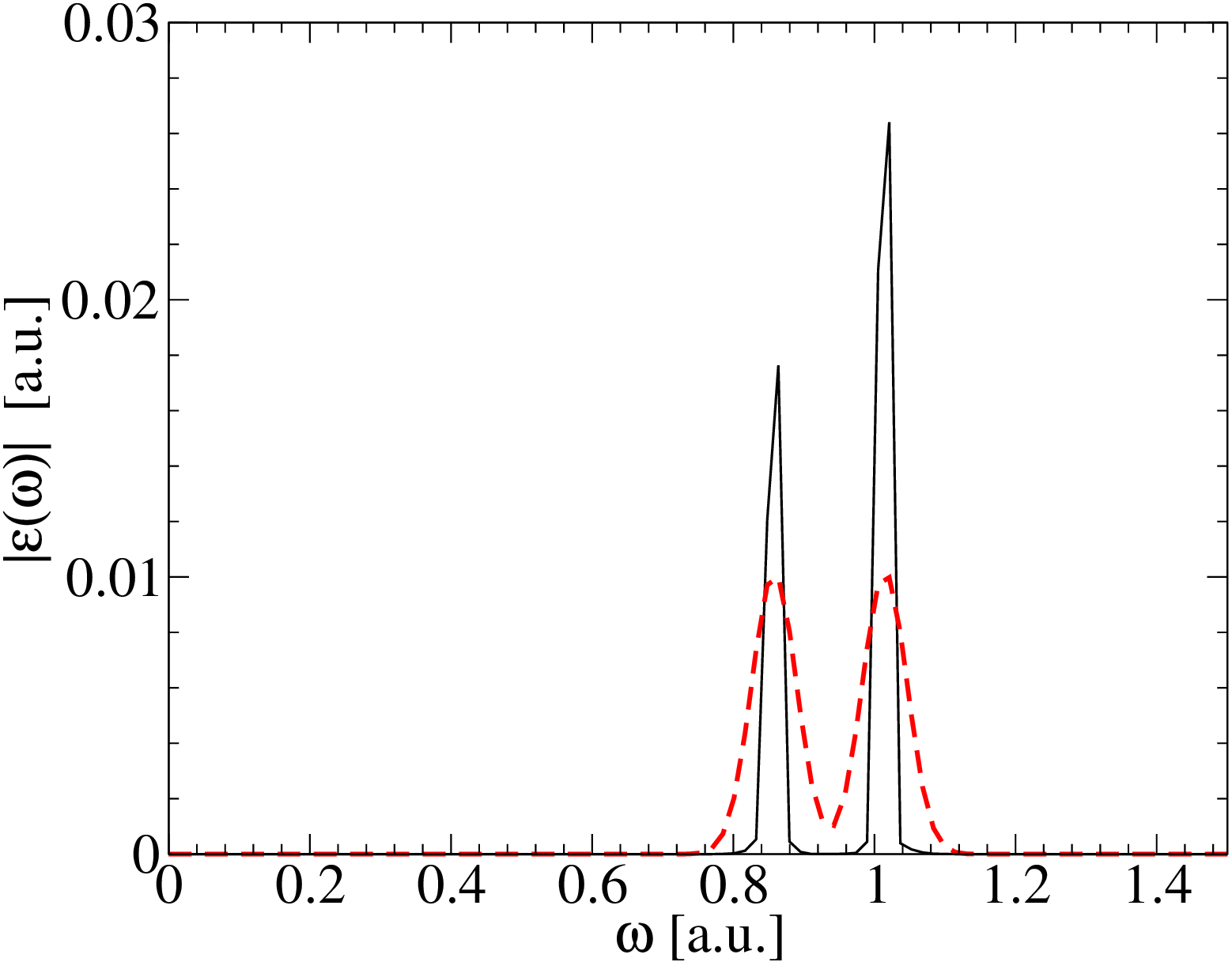}
}
\subfigure[]{
  \label{fig:field_FC031_3}
  \includegraphics*[width=.33\textwidth]{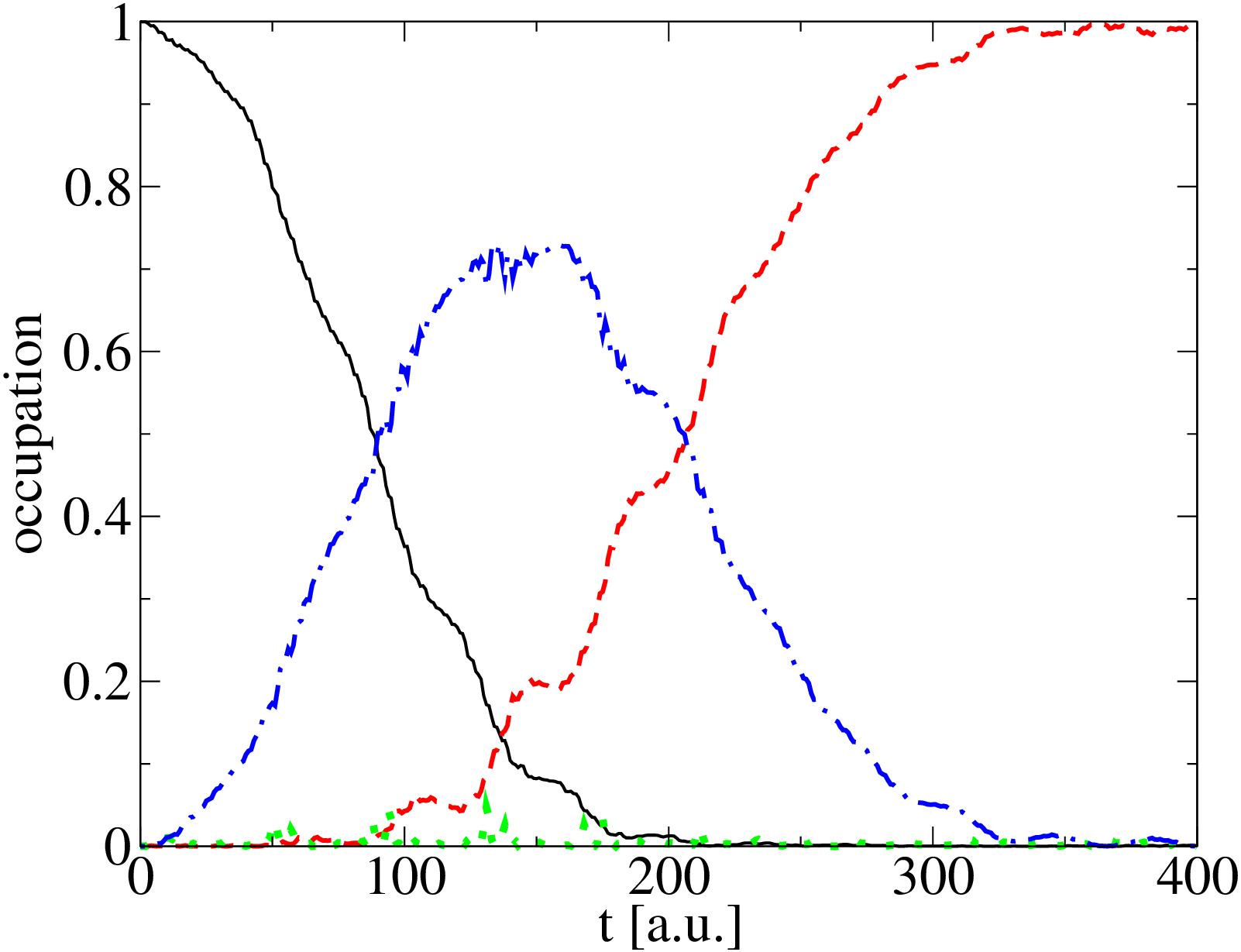}
}
\subfigure[]{
  \label{fig:field_FC031_4}
  \includegraphics*[width=.33\textwidth]{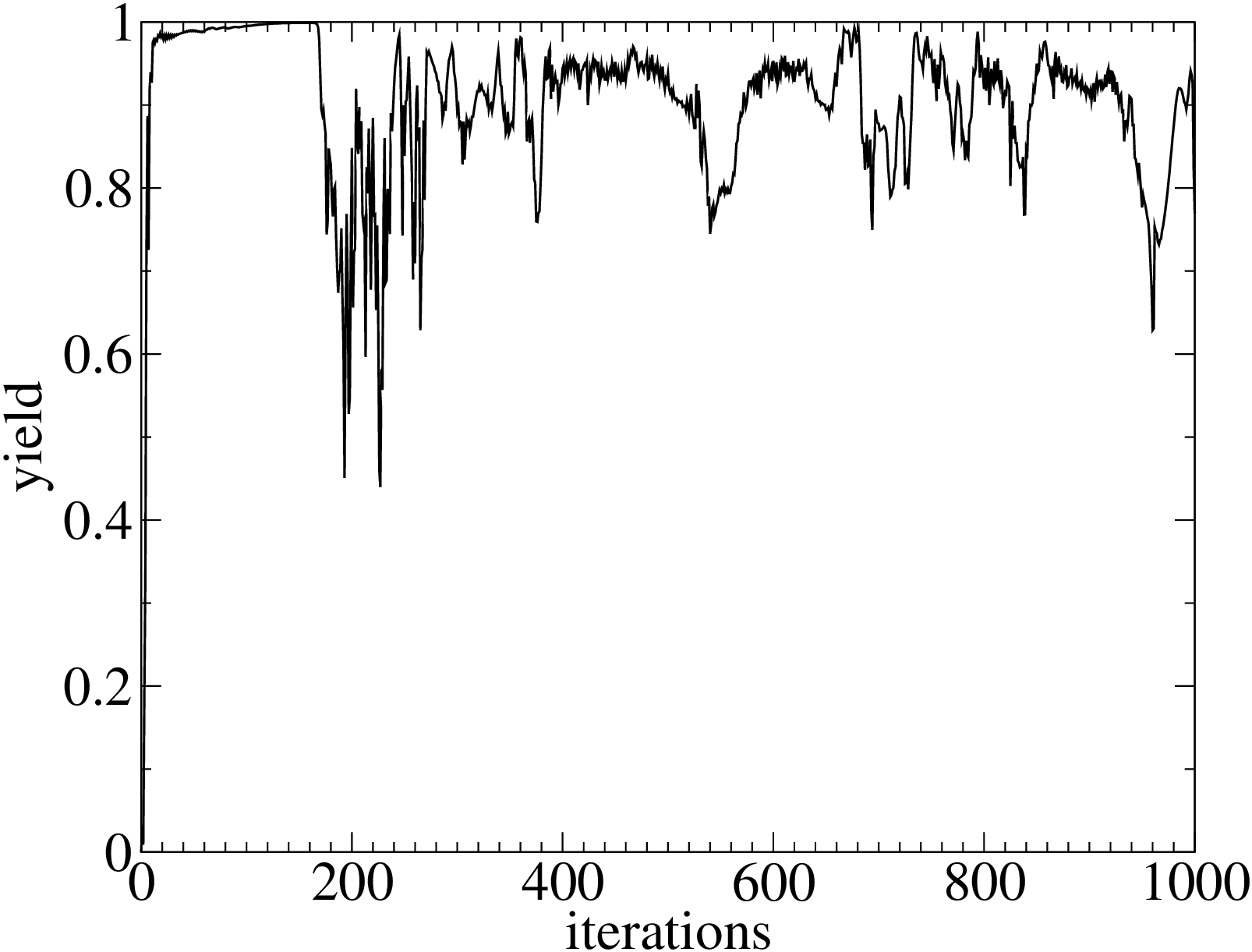}
}

\caption{We apply the optimization algorithm with a double gaussian window, allowing only frequencies around $\omega_{03}$ and $\omega_{31}$ and in addition set $E_0=0.320$. The optimized field is shown in graph (a). The filter function $f(\omega)$ (\dashed), scaled by $0.01$, is plotted together with the spectrum (b). The time-dependent occupation numbers (ground state (\full), first-excited state (\dashed) and second excited state (\dotted)), shown in (c), confirm that the transition process is performed mainly via the third excited state (\chain). Due to the strong restrictions the convergence behaviour happens to be more oscillatory, shown in graph (d). }
\label{fig:field_FC031}
\end{figure}

The results are shown in \fref{fig:field_FC031}. Like in the previous example, the high requirements on the laser field result in a rather erratic convergence (see \fref{fig:field_FC031_4}). The field, shown in \fref{fig:field_FC031_1}, produces a target state occupation of $99.89\%$ after $162$ iterations. The time-dependent occupation numbers (see \fref{fig:field_FC031_3}) show that the transition exactly happens in the desired way.

\subsubsection{Low frequency pulse}
\label{sec:lowfrequency}
\begin{figure}[!h]
\centering
\includegraphics*[width=.66\textwidth]{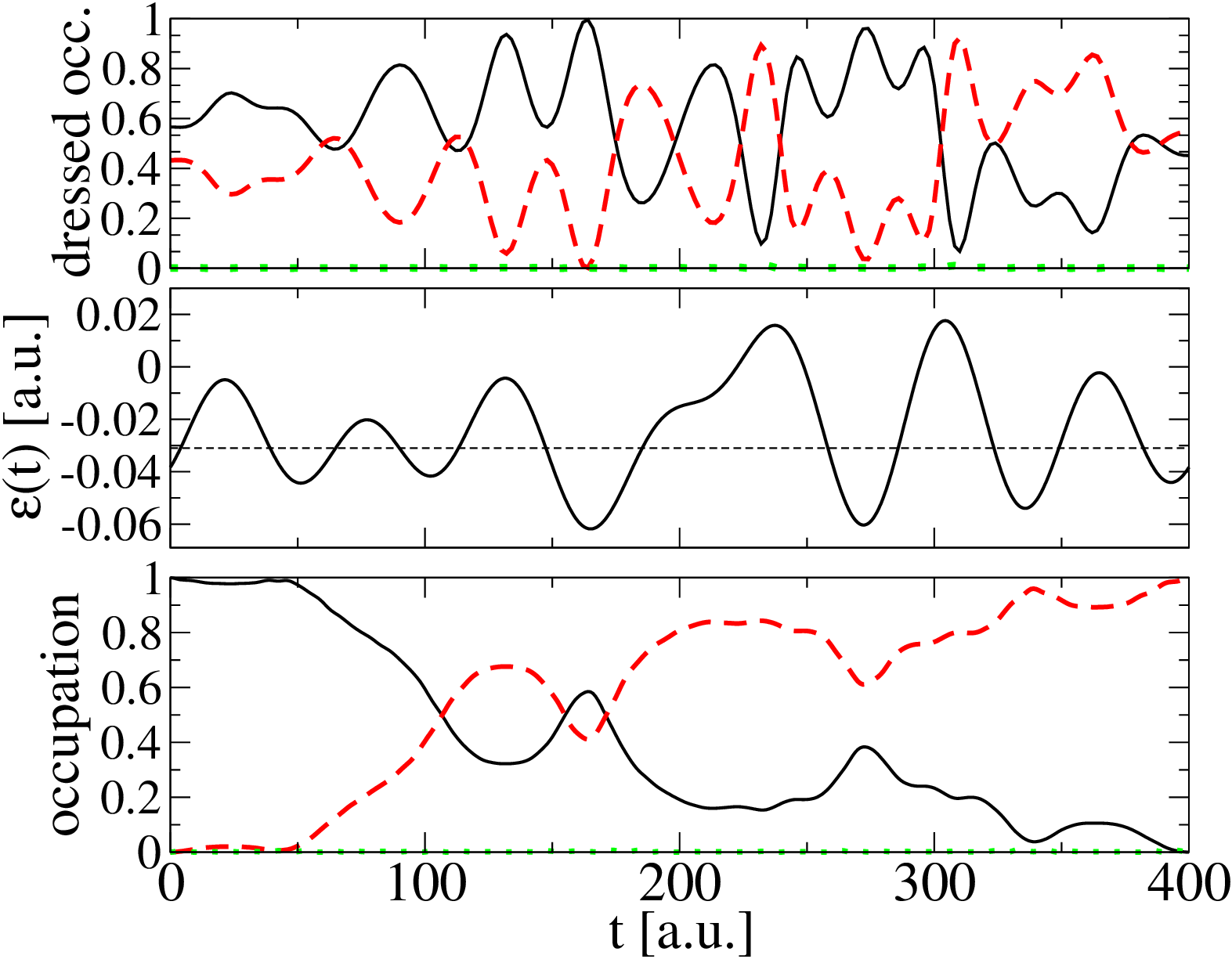}
\caption{In the lower panel we show the time-dependent occupation numbers of the ground-state (\full), the first excited state (\dashed) and the second excited state (\dotted).  The optimized laser field together with $\bar{\epsilon}=-0.031$ is shown in the middle panel. In the top panel we plot the absolute values (squared) of the projections onto the two lowest eigenfunctions of the Hamiltonian including the field $\bar{\epsilon}$.}
\label{fig:fieldS1}
\end{figure}

\begin{figure}[!h]
\centering 
\subfigure[]{
  \label{fig:field_S2}
  \includegraphics*[width=.33\textwidth]{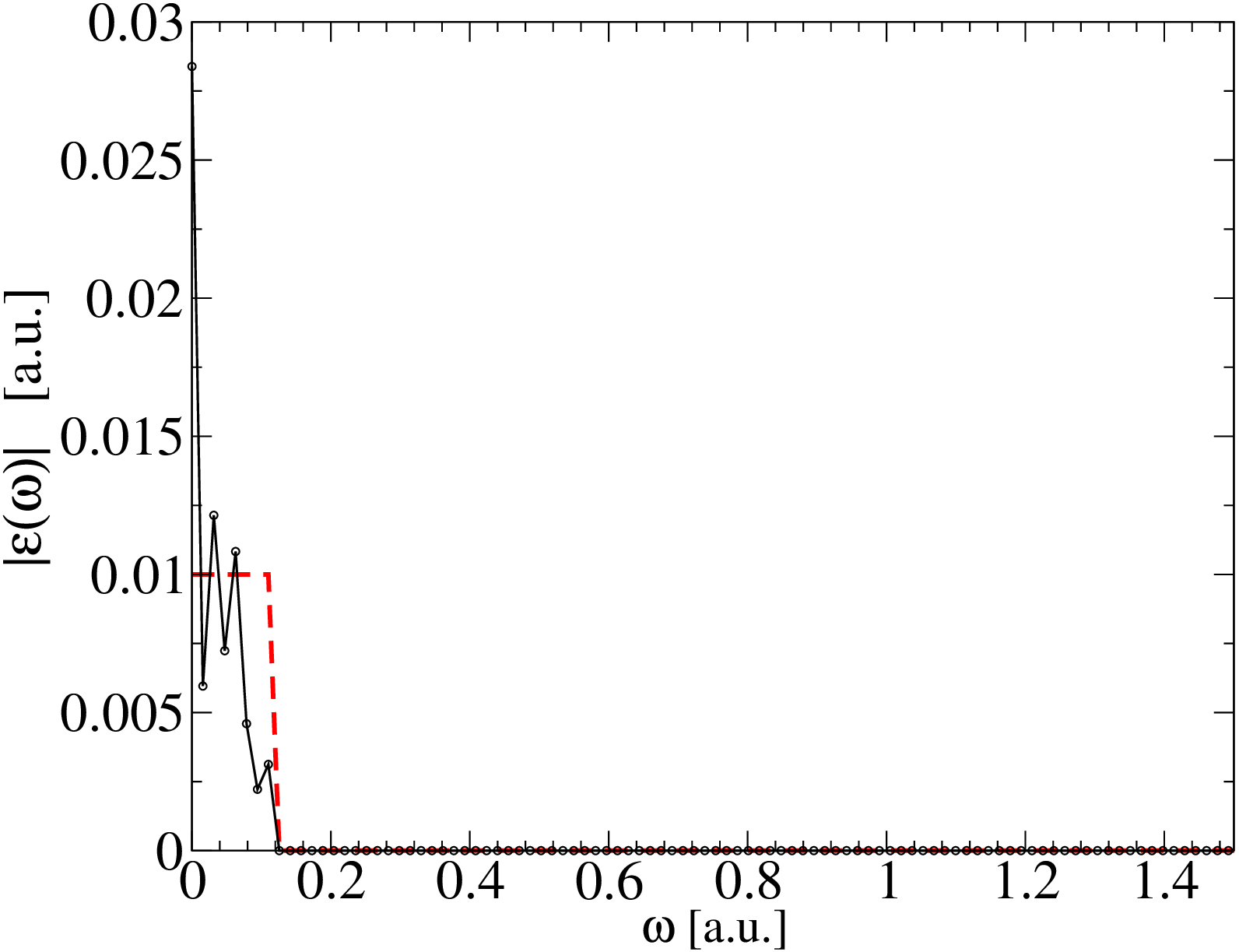}
}
\subfigure[]{
  \label{fig:field_S4}
  \includegraphics*[width=.33\textwidth]{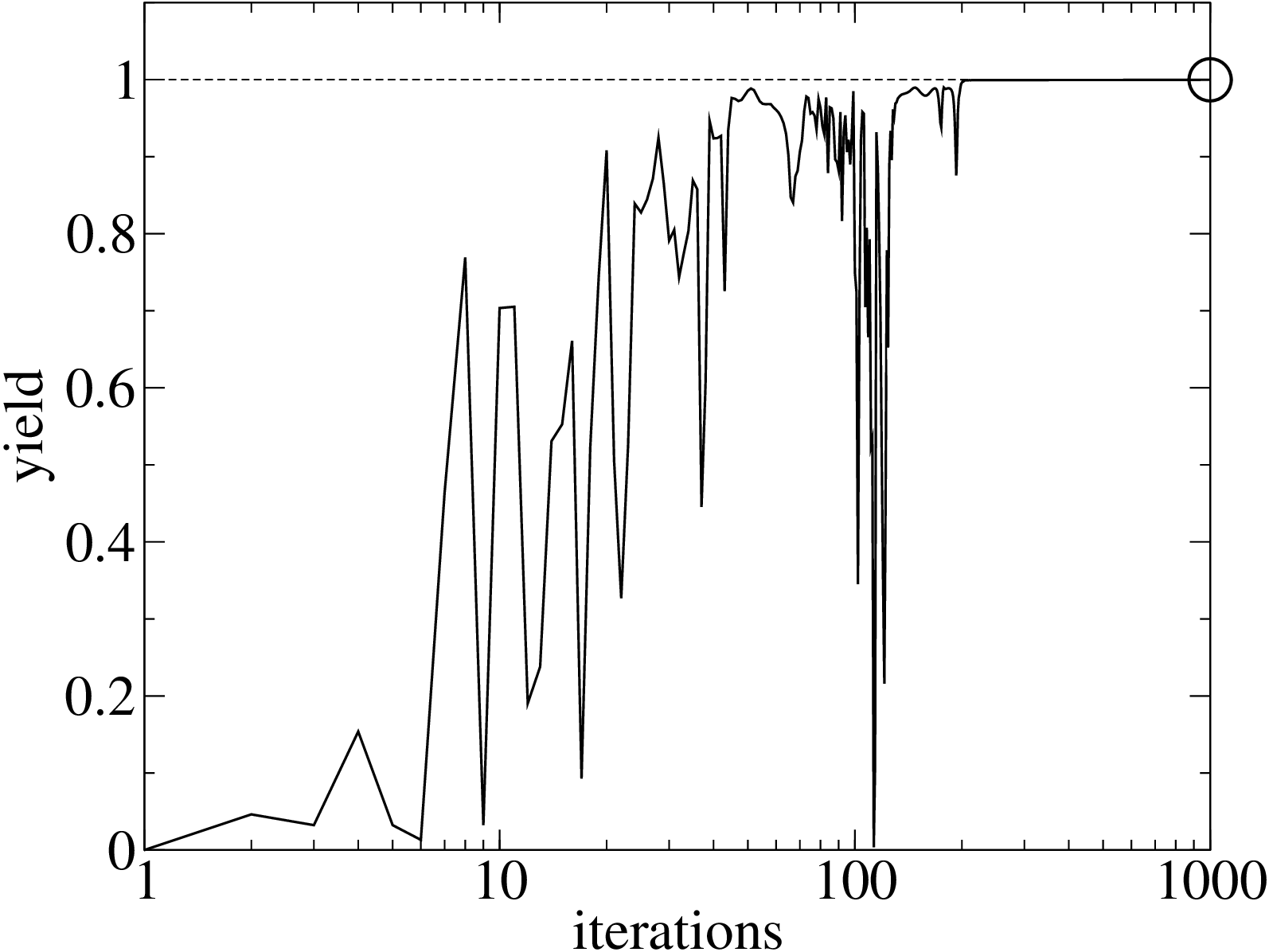}
}
\caption{ Graph (a) shows the spectrum of the laser field (\full) and the rectangular filter (\dashed) scaled by $0.01$. The convergence is shown in (b). The \opencircle indicates the iteration with the highest yield.}
\label{fig:field_S}
\end{figure}

%
Using a band filter, i.e. $f(\omega) = \theta(\omega - \omega_a)* \theta(\omega_b - \omega) +  \theta(-\omega_a - \omega)* \theta(\omega + \omega_b)$, with $\omega_a = 0.000$ and $\omega_b=0.120$ we can find a laser pulse resulting in high yields without allowing any resonance frequency in the laser spectrum. The allowed frequencies ($\omega \in [0.000,0.120]$) are smaller than the lowest excitation frequency $\omega_{01}$. The additional constraint on the fluence in the optimization is $E_0=0.400$. 
The convergence of this optimization is shown in \fref{fig:field_S4}. After $981$ iterations, we obtain a target state occupation of $99.93\%$. The spectrum of the optimized pulse, shown in \fref{fig:field_S2}, exhibits contributions of all allowed frequency components. In particular, the zero-frequency component is dominant, being almost three times larger than the other frequency contributions. In the time-domain the zero-frequency component corresponds to a bias of $\epsilon_{\mathrm{av}}=-0.028$ which is close to the value for which the potential becomes almost symmetric: $\bar{\epsilon}=-0.031$. The optimized pulse together with $\bar{\epsilon}$ are shown in the middle panel of \fref{fig:fieldS1}. 

The transfer process can be interpreted with the help of the following simplified picture: Assume the field would be almost static with $\epsilon(t) \approx \bar{\epsilon}$. Then, the initial state becomes a superposition of the dressed states
\begin{eqnarray}
\nonumber
\Psi(x,t=0) = \frac{1} {\sqrt{2}} \left( \varphi_0^{\mathrm{\bar{\epsilon}}} - \varphi_1^{\mathrm{\bar{\epsilon}}}\right) e^{i \theta_1}. 
\end{eqnarray}
Small perturbations of the laser field around $\bar{\epsilon}$ rearrange the phases of the superposition so that at the end of the pulse
\begin{eqnarray}
\nonumber
\Psi(x,T) =  \frac{1} {\sqrt{2}} \left(\varphi_0^{\mathrm{\bar{\epsilon}}} + \varphi_1^{\mathrm{\bar{\epsilon}}}\right)e^{i \theta_2}. 
\end{eqnarray}
This superposition is located in the right well which completes the transfer. Note, that the phases $\theta_1$, $\theta_2$ are irrelevant in this case.

The pulse we have obtained from the optimization is more difficult to explain since the oscillations around $\bar{\epsilon}$ are not small. To be able to analyse the transfer process in similar terms as discussed above, we have calculated the projections of the wave function $\Psi(x,t)$ onto the eigenfunctions of the Hamiltonian $\hat{H}^{\bar{\epsilon}}$ including the field $\bar{\epsilon}$. These ``dressed'' occupation numbers are shown in the upper panel of \fref{fig:fieldS1}.
In the simplified interpretation we have implicitly assumed a complete localization in the left (right) well for the initial (target) state. Since this is not true for the potential chosen here, the dressed occupation numbers deviate slightly from $0.5$, namely we have $|\langle \Psi(x,0) | \varphi_0^{\bar{\epsilon}} \rangle|^2 = 0.57$  and $|\langle \Psi(x,0) | \varphi_1^{\bar{\epsilon}} \rangle|^2 = 0.43$. At the end of the pulse we obtain the inverted occupation numbers, i.e. $|\langle \Psi(x,T) | \varphi_0^{\bar{\epsilon}} \rangle|^2 = 0.45$ and $|\langle \Psi(x,T) | \varphi_1^{\bar{\epsilon}} \rangle|^2 = 0.55$ which indicates the completed transfer (necessary condition). The laser pulse has also adjusted the phases of the expansion coefficients in the right way (sufficient condition): The relative phase difference of $\phi_0^{\mathrm{\bar{\epsilon}}}$ and $\phi_1^{\mathrm{\bar{\epsilon}}}$ between $t=0$ and $t=T$ was found to be $0.8*\pi$. This deviates from the simple picture where we would have expected a phase difference of $\pi$.

The transfer process described above has similarities with the one discovered in reference \cite{DKMS98} where the authors have used an asymmetric double well to model a hydrogen transfer reaction. 

%
\section{Conclusions}
%
We have presented a simple iterative scheme which allows for the optimization laser pulses under constraints on the spectrum of the laser and on its fluence. The scheme has been described in three different versions, one incorporating a given laser fluence, one restricting the spectrum of the laser pulse and the third one combining both constraints. Therefore, the scheme allows one to include realistic experimental constraints in the numerical optimization of laser pulses. 

To show that all three kinds of this scheme lead to high occupations of the target state we have applied them to drive the ground state of a 1D asymmetric double well potential to its first excited state. 

For all numerical tests we have obtained a high occupation $>99\%$ in the target state. In the case of a fixed fluence ($E_0=0.080$) we found a target state occupation of $99.90\%$ at the end of the pulse. Comparing the optimal laser pulses for different fluences shows that, for fluences larger than a certain critical value, target state occupations larger than $99\%$ can always be achieved. With increasing fluence the optimized pulses employ a growing number of transition processes.
Using spectral restrictions we are able to select between the different processes. We have calculated pulses that transfer the ground-state population to the target state only via the direct excitation process, explicitly without the direct process or via certain predefined intermediate levels. For the optimizations via an intermediate level we have additionally required a fixed laser fluence. 

That it is possible to achieve a very high target state occupation with laser fields not containing any of the excitation frequencies has been clearly demonstrated by the last example. The laser spectrum was allowed to have frequency components only lower than the lowest resonance frequency. In addition we required the fluence to be fixed. Like in the previous cases, the algorithm resulted in a laser pulse with a very high occupation of the target state.

The results obtained in this work, clearly demonstrate that, in general, there exists no unique optimal laser pulse to achieve a given control target and that selection within the set of optimal pulses is possible by adding constraints to the optimization.
With the methods presented here experimental constraints can be incorporated in the pulse optimization which makes the interpretation and analysis of the experimentally obtained laser pulses more reliable. The scheme allows one to study systematically the effects of different constraints on the target occupation and on the optimized laser field. Especially in the strong field regime, this leads to important insights in the various possible ways to achieve complete population transfer.

\appendix
\section*{Appendix}
\setcounter{section}{1}
%
\subsection{Ineffectiveness of post-constraints}
\label{app_post}
%
To show the ineffectiveness of filtering the pulse spectrum {\it after} the optimization we have taken the optimized pulse from section \ref{sec:fluence_2lev} and cut out the undesired frequencies. We then transform the pulse back to the time domain and propagate the time-dependent Schr\"odinger equation for the double well structure using this pulse.
Since the pulse has a lower fluence after this procedure we also rescale the pulse so that
\begin{equation}
\nonumber
 \int_0^T \!\! dt\,\, \epsilon^2(t) = 0.080. 
\end{equation}
We apply this procedure to two cases:
\begin{enumerate}
\item Case 1: Restricition to the direct process.\\
Filtering out all frequencies except $\omega \in [0.094,0.236]$, which corresponds to consider only the direct process $|0 \rangle \to |1\rangle$, results in a yield  of $8\%$ and after rescaling we obtain $44\%$. 
\item Case 2: Enforcing the indirect process. \\
By filtering out all frequencies except $\omega \in [0.503,0.833]$ we address only the indirect process $|0 \rangle \to |2\rangle \to |1\rangle$. Numerical propagation with the modified field results in a yield  of $5\%$ and $75\%$ after rescaling.
\end{enumerate}
Comparing these numbers to the high yields found in \sref{sec:direct} and \ref{sec:indirect021}, we clearly see that filtering after optimization is ineffective. It is far more powerful to use the filtering in the optimization process.

\subsection{Results from two-level system}
\label{app_2lev}
%
From the theory of two-level systems (or two-level atoms) \cite{AE75} we can extract a good estimate for an optimal pulse, if the direct transition is allowed in dipole approximation. The estimate is extremely good, if no more than two-levels contribute to the process. This is the case if the excitation spectrum is well separated and the laser pulse is in the weak response regime. We have chosen a pulse length $T=400$ which lies at the boundary of this  regime but since the excitation energies are far apart from each other, we expect the two-level system to be a good approximation. 
The optimal pulse for a two-level-system (within the rotating wave-approximation(RWA)) that transfers all population from the ground-state to the excited state is a simple sinusoidal oscillation \cite{SZ97,janphd}:
\begin{eqnarray}
\epsilon(t) = A \sin(\omega_{01}t)
\end{eqnarray} 
where $\omega_{01}$ is the resonance frequency and $A$ is the (optimal) amplitude given by:
\begin{eqnarray}
 A = \frac{\pi}{\mu_{01} T},
\end{eqnarray} 
with the dipole matrix element $\mu_{01}= \langle 0 | \hat{\mu} | 1 \rangle $ and $T$ the length of the pulse.
In our case we find $A = 0.02003$ and the corresponding fluence $E_0=0.0804$. Applying this pulse to the double well system, initially in the ground state, yields an occupation of $99.30\ \%$ of the first excited state. 
In \tref{tab:lev2_vs_opt} we compare the results obtained from this simple estimate with the optimal control solution fixed to the same fluence. The results show that the two-level estimate is very successful for long times, however for short pulse lengths where $T < 5*2 \pi / \omega_{01}$ it is not effective. This is due to the strength of the amplitude of the oscillation, it causes occupation also of the non-resonant levels.
\begin{table}[h]
\begin{indented} 
\item[]
\caption{\label{tab:lev2_vs_opt}Comparison of the yield $P=|\langle1 | \Psi(T)\rangle |^2$ obtained with the two-level (RWA) pulse estimate versus the optimal control result. Note, that the period of the oscillation with the resonance frequency $\omega_{01}$ is $T_p=40.08$~a.u.. }
\lineup
\begin{tabular}{@{}llll}
\br
 $T$   &  $P_{\mathrm{2level}}$  & $P_{\mathrm{opt}}$ & $E_0$    \\
\mr
  400 &  0.9930   & 0.9991  & 0.0804 \\    
  200 &  0.9042   & 0.9999  & 0.1608 \\
  100 &  0.1448   & 0.9999  & 0.3216 \\
  50  &  0.0199   & 0.9958  & 0.6407 \\
\br
\end{tabular}
\end{indented}
\end{table}
%
\ack
We would like to thank Stefan Kurth and Patrick Rinke for valuable discussions. This work was supported, in part, by the Deutsche Forschungsgemeinschaft, the EXC!TING Research and Training Network of the European Union and the NANOQUANTA Network of Excellence.

\section*{References}
\bibliographystyle{phpf}
\bibliography{paper}
\end{document}